\documentclass[12pt]{article}

\usepackage[T1]{fontenc}
\usepackage[latin1]{inputenc}
\usepackage{array}
\usepackage{subfigure}
\usepackage{graphicx}
\usepackage[left=2cm,top=2cm,bottom=2cm,right=2cm,nohead,nofoot]{geometry}



\begin{document}

\title{ Conditional probability based significance tests for sequential patterns in multi-neuronal spike trains }
\author{P.S. Sastry\footnote{Dept. Elec. Engg., Indian Institute of Science, 
Bangalore, India, email:sastry@ee.iisc.ernet.in},
and K.P. Unnikrishnan\footnote{General Motors R\&D Center, Warren, 
USA, email:k.p.unnikrishnan@gmail.com}}
\maketitle
\thispagestyle{empty}
\begin{abstract}
 In this paper we consider the problem of 
detecting statistically significant sequential patterns in multi-neuronal spike trains. 
These patterns are characterized by an ordered sequences of spikes from different neurons 
with specific delays between spikes. We have previously proposed a datamining scheme \cite{PSU2008}
to efficiently discover such patterns which 
 are frequent in the sense that the count of non-overlapping 
occurrences of the pattern in the data stream is above a threshold. Here we propose a method 
to determine the statistical significance of these repeating patterns and to set the thresholds 
automatically.  
The novelty of our approach is that we use a 
compound null hypothesis that includes not only models of independent neurons but also 
 models where neurons have weak dependencies. The strength of 
interaction among the neurons is represented in terms of certain pair-wise conditional 
probabilities. We specify our null hypothesis by putting an upper bound on all such 
conditional probabilities. 
We construct a 
probabilistic model that captures the counting process and use this 
 to calculate the mean and variance of the count for any pattern.  
Using this we  
 derive a test of significance for rejecting such a null 
hypothesis. This also allows us to rank-order different significant patterns. 
We illustrate the effectiveness of our approach using spike trains generated from a 
non-homogeneous Poisson model with embedded dependencies. 

\end{abstract}

\section{Introduction}
\label{sec:intro}

Analyzing spike trains from hundreds of neurons to find significant temporal patterns 
is an important current research problem \cite{Brown2004,SS2006,Pipa2008}. 
By using experimental techniques such as Micro Electrode Arrays or imaging of
neural currents, spike data can be recorded
simultaneously from many neurons \cite{Ikegaya2004,Potter2006}. Such multi-neuronal 
spike train data can 
now be routinely gathered {\em in vitro} from neural cultures or {\em in vivo} from 
brain slices, awake behaving animals and even humans. Such data would be a mixture 
of stochastic spiking activities of individual neurons as well as that due to correlated 
activity of groups of neurons due to interconnections, possibly triggered by external inputs. 
Automatically discovering patterns (regularities) in these spike trains can lead to better
understanding of how interconnected neurons act in a coordinated manner to generate 
specific functions. There has been much interest in techniques for analyzing the 
spike data so as to infer functional connectivity or 
 the functional relationships within the 
system that produced the spikes 
\cite{Abeles1988,Brillinger1992,Meister1996,Gerstein2004,Brown2004,Kass2005,Feber2007,sasaki2007,Ikegaya2004,Lee2004,SS2006,Pipa2008,FAHB2008}. 
In addition to contributing towards our 
knowledge of brain function, 
understanding of functional relations embedded in spike trains leads to many
applications, e.g., better brain-machine interfaces. Such an analysis
can also ultimately allow us to systematically address the 
question, "is there a neural code?".

In this paper, we consider the problem of discovering  {\em statistically 
significant} patterns from multi-neuronal spike train data. The patterns we 
consider here are ordered firing sequences by a group of neurons with 
specific time-lags or delays between successive neurons. 
Such a pattern (when it repeats many times) may denote a chain of 
triggering events and hence unearthing such patterns from spike data can 
help understand the underlying functional connectivity. For example, memory traces are 
probably embedded in such sequential activation of neurons and signals 
of this form have been found in hippocampal neurons \cite{Lee2002}. 
Such patterns of ordered firing sequences with fairly constant delays between 
successive neuronal firings have been observed in many experiments  and there is 
much interest in detecting such patterns and assessing their statistical 
significance. (See \cite{Abeles2001,Gerstein2004} and references therein). 

Here, we will call patterns of ordered firing sequences as {\em sequential  
 patterns}. 
Symbolically, we denote such a pattern as, e.g., 
$A \stackrel{T_1}{\rightarrow} B \stackrel{T_2}{\rightarrow} C$.  
This represents the pattern of ordered firing sequence of $A$ followed by $B$ 
followed by $C$ with a delay of $T_1$ time units between $A$ \& $B$ and 
$T_2$ time units between $B$ \& $C$. (We note here that within any occurrence of such 
a firing pattern, there could be spikes by other neurons).  
 Such a pattern of firings may occur repeatedly in the spike train data 
if, e.g., there is an excitatory influence of total delay $T_1$ from $A$ to $B$ and an 
excitatory influence of delay $T_2$ between $B$ and $C$. In general, the delays 
may not be exactly constant because synaptic transmission etc. could have some random 
variations. Hence, in our sequential patterns, we will allow the delays 
to be intervals  of small length. At the least, we can take the length of the 
interval as the time resolution in our measurements. In general, such patterns can 
involve more than three neurons. The {\em size} of a pattern is the number of 
neurons in it. Thus, the above example is that of a size 3 pattern or a 3-node 
pattern.  

 One of the main computational methods for detecting 
such patterns that repeat often enough,  is due to 
 Abeles and Gerstein \cite{Abeles1988}. This 
essentially consists of sliding the spike train of one neuron with respect to 
another and noting coincidences at specific delays. There are also some recent 
variations of this method \cite{Tetko2001a,Tetko2001b}. Most of the current 
methods for detecting such patterns essentially use correlations among time-shifted 
spike trains (and some statistics computed from the correlation counts),  
and these are computationally expensive when 
 detecting large-size (typically greater than 4)  
patterns \cite{Gerstein2004}. Another approach to detecting 
such ordered firing sequences is considered in \cite{Lee2004,SS2006} while analyzing 
recordings from hippocampal neurons. Given a specific ordering on a set of neurons, they 
look for longest sequences in the data that respect this order. This is similar to our 
sequential patterns which are somewhat more general because we can also specify different 
delays between consecutive elements of the pattern. 

 In this paper we use a method based on some 
temporal datamining techniques that we have recently proposed \cite{PSU2008}. 
This method can automatically detect all sequential patterns whose {\em frequency}  
in the data is above a (user-specified) threshold where {\em frequency} of the pattern 
is maximum number of non-overlapped occurrences\footnote{We define this notion more 
precisely in the next section} of the pattern in the spike data. The essence of 
this algorithm is that instead of trying to count all occurrences of the pattern 
in the data we count only certain well-defined subset of occurrences and this makes 
the process computationally efficient.  
The method is effective in detecting long patterns and it would detect only those 
patterns that repeat more than a given threshold. Also, the method can  
automatically decide on the most appropriate delays in each detected pattern by 
choosing from a set of possible delays supplied by the user. 
 (See \cite{PSU2008} for details).

The main contribution of this paper is a method for assessing the statistical 
significance of such sequential patterns. The objective is to have a method 
so that we will detect only those patterns that repeat often enough to be 
significant (and thus fix the thresholds for the data mining algorithm automatically). 
We tackle this issue in a classical hypothesis testing framework. 

There have been many approaches for assessing the significance of detected firing 
patterns \cite{Abeles2001,Gerstein2004,Lee2004,SS2006,FAHB2008}. 
In the current analytical approaches, one generally employs a Null hypothesis 
that the different spike trains are generated by independent processes. 
In most cases one assumes (possibly inhomogeneous) Bernoulli or  
Poisson processes. Then one can calculate the probability of observing the given 
number of repetitions of the pattern (or of any other statistic derived from such 
 counts) under the null hypothesis of independent processes and hence calculate 
a minimum number of repetitions needed to conclude that a pattern is significant 
 in the sense of being able to reject the null hypothesis. There are also  
 some  empirical approaches suggested for assessing significance 
\cite{date2001,Abeles2001,Gerstein2004,Nadasdy1999}. 
Here one creates many surrogate data streams from the 
experimentally observed data by perturbing the individual spikes while keeping 
certain statistics same and then assessing significance by noting whether or not 
the patterns are preserved in the surrogate data. There are many possibilities  
 for the perturbations to be imposed to generate  
surrogate data \cite{Gerstein2004}. In these empirical methods also, the implicit 
null hypothesis assumes independence.

The main motivation for the approach presented here is the following. If a  
 sequential pattern repeats often enough to be significant then 
one would like to think that there are strong influences among the 
neurons representing the pattern. However, different (detected) patterns  
may represent different 
levels or strengths of influences among their constituent neurons. Hence it would 
be nice to have a method of significance analysis that can rank order different 
(significant) patterns in terms some `strength of influence' among the neurons of 
the pattern. For this, here we propose that the strength of influence of $A$ on $B$ 
 is well represented by 
the conditional probability that $B$ will fire after some prescribed delay given that 
$A$ has fired. We then employ a composite null hypothesis specified through 
one parameter that denotes an upper bound on all such pairwise conditional 
probabilities. Using this we would be able to decide whether or not a given 
pattern is significant at various values for this parameter in the null 
hypothesis and thus be able to rank-order different patterns. 

There is an additional and important advantage of the above approach that we propose here. 
Our composite null hypothesis is such that any stochastic model for a set of 
spiking neurons would be in the null hypothesis if all the relevant 
pairwise conditional probabilities are below some bound. Since this bound  
 is a parameter that can be 
chosen by the user, the null hypothesis would include not only independent neuron 
models but also many models of interdependent neurons where the pair-wise 
influences among neurons are `weak'. Hence rejecting such a 
null hypothesis is more attractive than rejecting a null hypothesis of independence  
when we want to conclude that a significant pattern 
indicates `strong' interactions among the neurons. In this sense, the approach 
presented here extends the currently available methods for significance analysis. 

We analytically derive some bounds on the probability that our counting process would
come up with a given number of repetitions of the firing pattern if the data is
generated by any model that is contained in our compound null hypothesis. 
 As mentioned earlier, we use the number of non-overlapped occurrences 
of a pattern as our test statistic instead of the total number of repetitions 
and employ a temporal datamining algorithm for counting non-overlapped occurrences of sequential  
 patterns \cite{PSU2008}. 
 This makes our method attractive for discovering significant 
patterns involving large number of neurons also.  
We show the effectiveness of the method through extensive simulation experiments 
on synthetic spike train data obtained through a model of inter-dependent 
non-homogeneous Poisson processes.  

The rest of the paper is organized as follows. In section~\ref{sec:epi} we give a 
brief overview of temporal datamining and explain our algorithm for detecting 
sequential patterns whose frequency is above some threshold. The full 
details of the algorithm are available elsewhere \cite{PSU2008,archive-report} and we 
provide only some details which are relevant for understanding the statistical 
significance analysis which is presented in section~\ref{sec:stat}. 
In section~\ref{sec:simu}, 
we present some simulation results on synthetic spike train data to show the 
effectiveness of our method.  
We present results to show that our method 
 is  capable of ranking different patterns in terms of the synaptic 
efficacy of the connections. While we confine our attention in this paper to only 
sequential patterns, the statistical method we present can be generalized to 
handle other types of patterns. We briefly indicate this and conclude the paper 
with a discussion in section~\ref{sec:dis}.

\section{Frequent Episodes Framework for discovery of sequential patterns}
\label{sec:epi}

Temporal datamining is concerned with analyzing symbolic time series data to 
discover `interesting' patterns of temporal 
dependencies \cite{Srivats-survey2005,Morchen2007}. Recently we have proposed 
that some datamining techniques, based on the so called frequent episodes framework, 
are well suited for analyzing multi-neuronal spike train data \cite{PSU2008}. 
Many patterns of interest in spike data such as synchronous firings by groups 
of neurons, the sequential patterns explained in the previous section, 
and synfire chains which are a combination of synchrony and ordered firings, can be 
efficiently discovered from the data using these datamining techniques. While 
the algorithms are seen to be effective through simulations presented in 
\cite{PSU2008}, no statistical theory was presented there to address the question 
of whether the detected patterns are significant in any formal sense which is the main 
issue addressed in this 
paper.  In this 
section we first briefly outline the frequent episodes framework and then 
qualitatively describe this datamining technique for discovering frequently 
occurring sequential patterns. 

In the frequent episodes framework of temporal datamining. 
 the  data to be analyzed is a sequence of events
denoted by $\langle(E_{1},t_{1}),(E_{2},t_{2}),\ldots\rangle$ where $E_{i}$
represents an \textit{event type} and $t_{i}$ the \textit{time of occurrence} of
the $i^{th}$ event. $E_i$'s are drawn from a finite set of event types, $\zeta$.
The sequence is ordered with respect to time of occurrences of the events so
that, $t_i\le t_{i+1}$,  $\forall i$. The following is an
example event sequence containing 11 events with 5 event types.
\begin{small}
\begin{equation}
\langle(A,1),(B,3),(D,5),(A,5),(C,6),(A,10),(E,15),(B,15),(B,17),(C,18),(C,19)\rangle
\label{eq:data-seq}
\end{equation}
\end{small}

In multi-neuronal spike data, the event type of a spike event is the label of the 
neuron\footnote{or the electrode
number when we consider  multi-electrode array recordings without the
spike sorting step} which
generated the spike and the event has the associated time of occurrence.
The neurons in the ensemble under observation fire action potentials at
different times, that is, generate spike events. All these spike events are
strung together, in time order, to give a single long data sequence as needed for
frequent episode discovery. It may be noted that there can be more than one event 
with the same time because two neurons can spike at the same time.  

The temporal patterns that we wish to discover in this
framework are called episodes. In general, episodes are partially ordered 
sets of event types. Here we are only interested in {\em serial episodes} which 
are totally ordered. 

A \textit{serial episode} is an ordered tuple of event types. For example,
$(A\rightarrow B\rightarrow C)$ is a {\em 3-node} serial episode. (We also say 
that the size of this episode is 3). The arrows in this
notation indicate the order of the events. Such an episode is said to
\textit{occur} in an event sequence if there are  corresponding events in the
prescribed order in the data sequence.
In sequence (\ref{eq:data-seq}), the events
\{${(A,1),(B,3),(C,6)}$\} constitute an occurrence of the
serial episode $(A\rightarrow B \rightarrow C)$ while
the events \{$(B,3), (C,6), (A,10)$\} do not.
We note here that occurrence of an episode does not
require the associated event types to occur consecutively;
there can be other intervening events between them.

 In the multi-neuronal data, if neuron $A$ makes
neuron $B$ to fire, then, we expect to see $B$ following $A$ often. However, in
different occurrences of such a substring, there may be different number of
other spikes between $A$ and $B$ because many other neurons may also be spiking
during this time. Thus, the episode structure allows us to unearth patterns
in the presence of such noise in spike data.

The objective in frequent episode discovery is to detect {\em all} frequent episodes 
(of different lengths) from the data. 
A {\em frequent episode} is one whose {\em frequency} exceeds a 
 (user specified) {\em frequency threshold}.
The frequency of an episode can be defined in many ways. 
It is intended
to capture some measure of how often an episode occurs in an event
sequence. One chooses a measure of frequency so that frequent episode discovery is
computationally efficient and, at the same time, higher frequency would imply that
an episode is occurring often.

Here, we define frequency of an episode as the maximum number of non-overlapped  
occurrences  of the episode in the data stream. 
Two occurrences of an episode  are
said to be \textit{non-overlapped} if no event associated with one occurrence appears in
between the events associated with the other. A set of occurrences is 
said to be non-overlapped if every pair of occurrences in it 
are non-overlapped. 
In our example sequence (\ref{eq:data-seq}), there are two non-overlapped 
 occurrences of $A \rightarrow B \rightarrow C$ given by the events: 
$( (A,1),(B,3),(C,6))$  and $( (A,10),(B,15),(C,18))$. Note that there are three 
distinct occurrences of this episode in the data sequence though we can have only a 
maximum of two non-overlapped occurrences. We also note that 
if we take the occurrence of the episode given by 
$((A,1),(B,15),(C,18))$, then there is no other occurrence that is non-overlapped with this 
occurrence. That is why we define the frequency to be the maximum number of non-overlapped 
occurrences. 

This definition of frequency results in very efficient
counting algorithms with some interesting theoretical
properties \cite{Srivats2005,Srivats-kdd07}.  
In addition, in the context of our application, counting non-overlapped occurrences seems 
natural because we would then be looking at chains that happen at
different times again and again. 

In analyzing neuronal spike data, it is useful to consider
methods,
where, while counting the frequency, we include only those occurrences which
satisfy some additional temporal constraints. Here we are interested in what we 
call inter-event time constraint which is specified by giving an interval of 
the form $(T_{low}, T_{high}]$. The constraint  
 requires that the difference between the times of
 every pair of successive events in any occurrence of a serial episode
should be in this interval.  In  general, we may have
different time intervals for different pairs of events in each serial episode. 
As is easy to see, a serial episode with inter-event time constraints corresponds 
to what we called a {\em sequential pattern} in the previous section. These are 
the temporal patterns of interest in this paper. 

\begin{figure}
\centering
\includegraphics[scale=1.0]{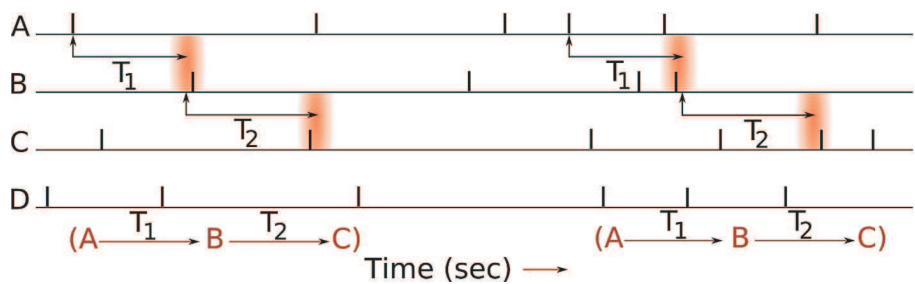}
\caption{ A schematic showing two occurrences of the sequential pattern 
$A\stackrel{T_1}{\rightarrow} B \stackrel{T_2}{\rightarrow} C$ in the spike trains from 
neurons $A, B, C, D$. A small interval (usually 1 ms) is shown around the second and third spike 
to indicate possible variation in the delay. Note that within the duration of one occurrence 
of the pattern there may be other intervening spikes (from any of the neurons). }
\label{fig:ser-epi-fig}
\end{figure}

The inter-event time constraint allows us to take care of delays involved in the process 
of one neuron influencing another through a synapse. Suppose 
neuron $A$ is connected to neuron $B$ which, in turn, is connected to neuron $C$,  
through excitatory connections with delays $T_1$ and $T_2$ respectively. Then, we should be 
counting only those occurrences of the episode $A\rightarrow B \rightarrow C$, where the 
inter-event times satisfy the delay constraint. This would be the sequential pattern 
$A\stackrel{T_1}{\rightarrow} B \stackrel{T_2}{\rightarrow} C$. In general, the inter-event 
constraint could be an interval. 
Occurrences of such serial episodes with inter-event constraints in spike data are shown schematically 
in fig.~\ref{fig:ser-epi-fig}

In any occurrence of the episode or 
sequential pattern, we call the difference between the times of the first and last 
events as its {\em span}. The span would be the total of all the delays. 
If, in the above episode, 
the span of all occurrences would be $T_1 + T_2$ and hence we may call it the span 
of the episode. If the inter-event time 
constraints are intervals then the span of different occurrences could be different.   

There are efficient algorithms for discovering all frequent serial episodes with specified 
inter-event constraints \cite{PSU2008}.  
That is, for discovering all episodes whose frequency (which 
is the number of non-overlapped occurrences of the episode) is above a given threshold.  

{\em Conceptually, the algorithm does the following}. 
 Suppose, we are operating at a time resolution of $\Delta T$. (That is, 
 the times of  of events 
or spikes are recorded to a resolution of $\Delta T$). 
Then we discretize the time axis into intervals of length $\Delta T$.
 Then, for each episode whose frequency we want to find we 
do the following. Suppose the episode is the one mentioned above. 
We start with time instant 1. We check to see whether there is an 
occurrence of the episode starting from the current instant. For this, we need an $A$ at that 
time instant and then we need a $B$ and a $C$ within appropriate time windows. If there 
are such $B$ and $C$, then we take the earliest  of the $B$ and $C$ to satisfy the 
time constraints, increment the counter for the episode and start looking for the occurrence 
again starting with the next time instant (after $C$). On the other hand, 
if we can not find such an occurrence (either because $A$ does not occur at the 
current time instant or because there are no $B$ or $C$ at appropriate times following 
$A$), then we move by one time instant and start the search again. 

The actual search process would be very inefficient if 
implemented as described above. The algorithm itself does the search in a much more 
efficient manner.  There are two issues that the algorithm needs to address. Since, 
a priori, we do not know what patterns to look for, we need to make a reasonable 
list of candidate patterns and then obtain their frequencies so as to output only 
those patterns whose frequency exceeds the preset threshold. The second issue is that 
in obtaining frequencies, the algorithm is required to count the frequencies of 
not one but a set of candidates in one pass through the data  
 and we need to do this efficiently. 
In generating the candidates, we need to tackle the combinatorial explosion 
because all possible serial episodes of a given size increases exponentially 
with the size.  
This is tackled using an iterative procedure that is popular 
in datamining. To understand this, consider our example 3-node pattern 
 $A\stackrel{T_1}{\rightarrow} B \stackrel{T_2}{\rightarrow} C$. This can not 
be frequent unless certain 2-node {\em subepisodes} of this, namely, 
$A\stackrel{T_1}{\rightarrow} B$ and  $B \stackrel{T_2}{\rightarrow} C$ are 
frequent. (This is because any two non-overlapped occurrences of the 3-node pattern 
also gives us two non-overlapped occurrences of the two 2-node patterns mentioned 
above).  Thus, we should allow this 3-node episode to be a candidate only if 
the appropriate 2-node episodes are already found to be frequent. Based on this 
idea, we have the following structure for the algorithm. 
 We first get frequent 1-node episodes which are then used to make candidate 2-node 
episodes. Then, by one more pass over data, we find frequent 2-node episodes which are 
then used to make candidate 3-node episodes and so on.  
Such a technique is quite effective in controlling combinatorial explosion and 
 the number of candidates comes down drastically as the size increases. This is because, 
as the size increases, many of the combinatorially possible serial episodes of that 
size would not be frequent.  This allows 
the algorithm to find large size frequent episodes efficiently. 
 At each stage of this process, we count frequencies 
of not one but a whole set of candidate episodes 
(of a given size) through one sequential pass over 
the data. We do not actually traverse the time axis in time ticks once for each 
pattern  whose occurrences we want to count. We traverse the 
time-ordered data stream. As we traverse the data we remember enough from the data stream to 
correctly take care of all the occurrence possibilities of all episodes in the candidate set 
and thus compute all the frequent episodes of a given size through one pass over the data.  
The complete details of the algorithm are available in \cite{PSU2008}. 


\section{Statistical Significance of Discovered Episodes or Serial Firing Patterns}
\label{sec:stat}

In this section we address the issue of the statistical significance of the 
sequential patterns discovered by our algorithm. 
The question is when are the discovered  
episodes significant, or, equivalently, what frequency threshold should we choose 
so that all discovered frequent episodes would be statistically significant. 

To answer this question we follow a classical hypothesis testing framework. 
Intuitively we want significant sequential patterns to represent a 
chain of strong interactions among those neurons. 
So, we have to essentially choose a {\em null hypothesis} that asserts
that there is no `{\em structure}' or `{\em strong influences}' 
in the system of neurons generating the data. Also, as mentioned earlier, we want the 
null hypothesis to contain a parameter that allows us to specify what we mean 
by saying that the influence one neuron has on another is not `strong'.

For this, we capture the strength of interactions among 
the neurons in terms of conditional probabilities. Let  
$e_s(A, B, T)$ denote the conditional probability that $B$ fires in a time 
interval $[T, T + \Delta T]$ given that $A$ fired at time zero.  $\Delta T$ is 
 essentially the time resolution at which we operate. 
(For example, $\Delta T$ = 1ms). Thus, 
$e_s(A, B, T)$ is essentially, the conditional probability 
that $B$ fires $T$ time units after $A$.\footnote{For the analysis,  
we think of the delay, $T$, as a constant. However, in practice our method 
can easily take care of the case where  
the actual delay is uniformly distributed over a small interval with 
$T$ as its expected value.}  If there is a strong excitatory  
synapse of delay $T$ between $A$ and $B$, then this conditional probability would be 
high. On the other hand if $A$ and $B$ are independent, then, this 
conditional probability is the same as the unconditional probability of $B$ firing 
in an interval of length $\Delta T$.  We denote the 
(unconditional) probability that a neuron, $A$, 
fires in any interval of length $\Delta T$ by $\rho_A$. 
(For example, if we take $\Delta T = 1ms$ and that
the average firing rate of $B$ is 20Hz, then $\rho_B$ would be
about 0.02).  

The main assumption we make is that the conditional probability $e_s(A,B,T)$ is 
not a function of time. That is, the conditional probability of $B$ firing at least 
once in an interval $[t+T, \ t+T+\Delta T]$ given that $A$ has fired at $t$ 
 is same for all $t$ within the 
time window of the observations (data stream) that we are analyzing. We think this  
is a reasonable assumption and some recent analysis of spike trains from 
 neural cultures suggests that such an assumption is justified  \cite{Feber2007}.
Note that this assumption does not
mean we are assuming that the firing rates of neurons are not time-varying. As a matter
of fact, one of the main mechanisms by which this conditional probability is realized
is by having a spike from $A$ affect the rate of firing by $B$ for a short duration
of time. Thus, the neurons would be having time-varying firing rates even when the
conditional probability is not time-varying. Essentially, the constancy of
$e(A,B,T)$ would only mean that every time $A$ spikes, it has the same chance of
eliciting a spike from $B$ after a delay  of $T$. Thus our assumption only
means that there is no appreciable change in synaptic efficacies during the period
 in which the data being analyzed  is gathered. 

The intuitive idea behind our null hypothesis is that the conditional probability 
$e_s(A,B,T)$ is a good indicator of the `strength of interaction' between $A$ and 
$B$. For inferring functional connectivity from repeating sequential patterns, 
the constancy of delays (between spikes by successive neurons) in multiple 
repetitions is important. That is why we defined the conditional probability with 
respect to a specific delay. Now, an assertion that the interactions 
among neurons is `weak' can 
be formalized in terms of an upper bound on all such conditional probabilities. 
We formulate our composite null hypothesis as follows. 

{\em Our composite null hypothesis includes all models of interacting neurons for which 
we have $e_s(x, y, T) < e_0$ for all pairs of neurons $x, y$ and for a set of 
specified delays $T$, where $e_0$ is a fixed user-chosen number in 
the interval $(0, \ 1)$.}
 
Thus all models of inter-dependent neurons where the probability of $A$ causing 
$B$ to fire (after a  delay) is less that $e_0$,  
would be in our Null hypothesis. The actual mechanism by which spikes from $A$ affect 
the  firing by $B$ is immaterial to us. Whatever may be this mechanism 
of interaction, if the resulting conditional probability is less than $e_0$, then that 
model of interacting neurons would be in our null hypothesis.\footnote{We note here 
 that this conditional probability is well defined whether or not the two 
neurons are directly connected. If they are directly connected then 
$T$ could be taken as a typical delay involved in the process; otherwise it can be taken as some 
integral multiple of such delays. 
 In any case, our interest is in deciding on the significance 
of sequential patterns with some given values for $T$.}  The user specified number, $e_0$, 
 formalizes what we mean by interaction among neurons is strong. If $A$ and $B$ are 
independent then this conditional probability is same as $\rho_B$. As mentioned 
earlier, if $\Delta T = 1ms$ and average firing rate for $B$ as 20 Hz, 
then $\rho_B=0.02$. So, if we choose $e_0=0.4$, it means that we agree to call the 
influence as strong if the conditional probability is 20 times what it would be if 
the neurons are independent.  By having different values for $e_0$ in the null 
hypothesis, we can ask what patterns are significant at what value of $e_0$ and thus 
rank-order patterns. 

Now if we are able to reject this Null hypothesis then it is 
reasonable to assert that the episode(s) 
discovered would  indicate `strong' interactions among the appropriate neurons. 
The `strength' of interaction is essentially chosen by us in terms of the 
bound $e_0$ on the conditional probability in our null hypothesis.

We now present a method for 
 bounding the probability that the frequency (number 
of non-overlapped occurrences) of a given serial episode with inter-event constraints 
is more than a given threshold under the null hypothesis. To do this, we first compute 
the expectation and variance (under the null hypothesis) 
of the random variable representing the  number of 
non-overlapped  occurrences of a serial episode with inter-event constraints  
 by using the following stochastic model.

 Let $\{X_i, \ i=1,2,\ldots \}$ be {\em iid} random 
variables with distribution given by
\begin{eqnarray}
P[X_i = T] &=&  p \nonumber \\
P[X_i = 1] &=& 1 - p 
\label{eq:xi}
\end{eqnarray}
where $T$ is a fixed constant (and $T>1$). Let $N$ be a random variable defined by 
\begin{equation}
N = \min\: \{ n \: : \: \sum_{i=1}^n X_i \geq L\}
\label{eq:N}
\end{equation}
where $L$ is a fixed constant. 

Let the random variable $Z$ denote the number of $X_i$'s out of the first $N$ 
which have value $T$. Define the random variable $M$ by
\begin{eqnarray}
M &=& Z \ \ \mbox{~if~} \ \  \sum_{i=1}^N X_i = L \nonumber \\
M &=& Z - 1 \ \ \mbox{~if~} \ \  \sum_{i=1}^N X_i > L 
\label{eq:M}
\end{eqnarray}

All the random variables, $N, Z, M$ depend on the parameters $L, T, p$. When it 
is important to show this dependency we write $M(L, T, p)$ and so on. 

Now we will argue that $M(L,T,p)$ is the random variable representing 
 the number of non-overlapped occurrences of an episode where $T$ is the span 
 (or sum of all delays) of the episode and $L$ is the length of data (in 
terms of time duration). We would fix $p$ based on the bound $e_0$ in our null hypothesis 
as explained below. 

\begin{figure}
\centering
\includegraphics[scale=0.5,clip]{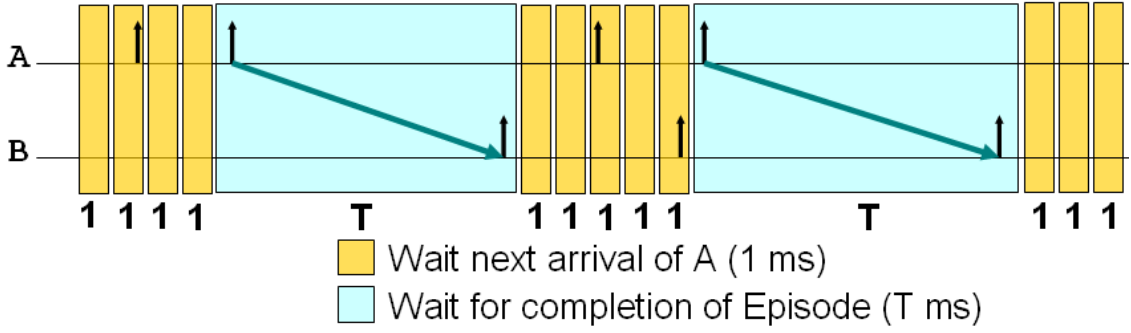}
\caption{ A schematic of the counting process for non-overlapped occurrences 
of the episode $A\stackrel{T}{\rightarrow}B$ superimposed on the spike trains 
from neurons $A$ and $B$. In the yellow region there are no occurrences of the pattern 
starting with that time instant and the counting scheme moves forward by one time step. 
In the blue region there is an occurrence and the counting process moves by $T$ time steps. 
The random variables $X_i$, defined by eq.~(\ref{eq:xi}), capture the evolution of the 
counting process}  
\label{fig:counting}
\end{figure}

Consider an episode $A \stackrel{T}{\rightarrow} B$ with an inter-event 
 time constraint (or  delay)  
of  $T$. Now, the sequence $X_i$ essentially captures the counting 
process of our algorithm. A schematic of the counting process (as relevant for this discussion) 
is shown in fig.~\ref{fig:counting}. As explained earlier, the algorithm can be viewed 
as traversing a {\em discretized} time axis in steps of $\Delta T$, 
looking for an occurrence of the episode 
starting at each time instant.  At each time instant (which, on the discretized 
time axis corresponds to an interval of length $\Delta T$),  
let $q_1$ denote the probability of spiking by $A$ and let $q_2$ denote the 
conditional probability that $B$ generates a spike  $T$ instants later 
given that $A$ has spiked now. In terms of our earlier notation, $q_1 = \rho_A$, 
$q_2 = e_s(A,B,T)$. 
Thus, at any instant, $q_1q_2$ denotes the probability 
of occurrence of the episode starting at that instant. Now, in eq.(\ref{eq:xi}) let 
$p=q_1q_2 (= \rho_A  e_s(A, B, T))$. Then $p$ represents the probability that this 
episode occurs starting with any given time instant.\footnote{We note here that we actually 
do not know this $p$ because we do not know the exact value for $e_s(A,B,T)$. But 
finally we would bound the relevant probability by using $e_0$ to bound 
$e_s(A,B,T)$.}  
Let $L$ in eq.(\ref{eq:N}) denote the data length (in time units). Then the sequence, 
$X_1, X_2, \ldots, X_N$,  represents our counting process. If, at the 
first instant there is an occurrence of the episode starting at that instant then 
we advance by $T$ units on the time-axis and then look for another occurrence (since 
we are counting non-overlapped occurrences); if there is no occurrence starting at the 
first instant then we advance by one unit and look for an occurrence. 
Also,  whether or not there is 
an occurrence starting from the current instant is independent of how many occurrences are 
completed before the current instant (because we are counting only non-overlapped 
occurrences). 
So, the counting process is well captured by accumulating the $X_i$'s defined above 
till we reach the end of data. Hence $N$ captures the number of such $X_i$ that 
we accumulate because $L$ is the data length in terms of time. 
 Since $X_i$ take values $1$ or $T$, the 
only way $\sum X_i$ exceeds $L$ is if the last $X_i$ takes value $T$ which in turn 
implies that when we reached end of data we have a partial occurrence of the episode. In 
this case the total number of completed  occurrences is one less than the number 
of $X_i$ (out of $N$) that take value $T$. If the last $X_i$ has taken value 1 (and 
hence the sum is equal to $L$) then the number of completed  occurrences is equal to 
the number of $X_i$ that take value $T$. Now, it is clear that $M$ is the 
number of non-overlapped occurrences counted. 

It is easy to see that the model captures counting of episodes of arbitrary length also. 
For example, if our episode is 
$A \stackrel{T_1}{\rightarrow} B \stackrel{T_2}{\rightarrow} C$ then $T$ is eq.(\ref{eq:xi}) 
would be $T_1+T_2$ and $p$ would be $\rho_A e_s(A,B,T_1) e_s(B,C,T_2)$.\footnote{Here we are 
 assuming that firing of $C$ after a delay of $T_2$ from $B$, conditioned on 
firing of $B$,  is conditionally 
independent of earlier firing of $A$. Since our objective is 
to unearth significant triggering chains, this is a reasonable assumption. Also, this allows 
us to capture the null hypothesis with a single parameter $e_0$. We discuss this further 
in section~\ref{sec:dis}.}
 Suppose in a $n$-node episode the conditional probability 
of $j^{\mbox{th}}$ neuron firing (after the prescribed delay) given that the previous 
one has fired,   is equal to $e_s^j$. Let the successive delays be 
$T_i$. Let the (unconditional) probability of the first neuron (of the episode) 
firing at any instant (that is, in any interval of length $\Delta T$) is $\rho$. 
Then we will take (for the $n$-node episode) $p=\rho \Pi_{j=2}^{n}(e_s^j)$ and $T= \sum T_i$. 


\subsection{ Mean and Variance of $M(L, T, p)$} 

Now, we first derive some recurrence relations to calculate the mean and variance of 
$M(L, T, p)$ for a given episode. Fixing an episode fixes the value of $p$ and $T$. 
Let $F(L,T,p) = E\:M(L,T,p)$ where $E$ denotes expectation. 
We can derive a recurrence relation for $F$ as follows.

\begin{eqnarray}
E\:M(L,T,p) & = & E\: \left[\; E\: [ M(L,T,p) \; | \; X_1 ] \; \right] \nonumber \\
 & = & E\: [M(L,T,p) \; | \; X_1 =1] (1-p) \; + \: E[M(L,T,p) \; | \; X_1 \neq 1] p \nonumber \\
& = & (1-p) E\:[M(L-1, T, p)] \: + \: p ( 1 \: + \: E[M(L-T, T,p)] ) \nonumber \\
& & 
\end{eqnarray}

In words what this says is: if the first $X_i$ is $1$ (which happens with probability $1-p$), 
then the expected number of occurrences is same as those in data of length $L-1$; on 
the other hand, if first $X_i$ is not $1$ (which happens with probability $p$) then 
the expected number of occurrences are 1 plus the expected number of occurrences in 
data of length $L-T$. 

Hence our recurrence relation is:
\begin{equation}
F(L,T,p) = (1-p) F(L-1, T, p) + p ( 1 + F(L-T, L, p))
\label{eq:rec1}
\end{equation}
The boundary conditions for this recurrence are:
\begin{equation}
F(x,y,p) = 0, \ \ \mbox{if} \ \ x < y \ \ \mbox{and} \ \  \forall p.
\label{eq:bd-cn}
\end{equation}

Let $G(L,T,p) = E[M^2(L,T,p)]$. That is $G(L,T,p)$ is the second moment of $M(L,T,p)$. 
Using the 
same idea as in case of $F$ we can derive recurrence relation for $G$ as follows. 
\begin{eqnarray}
E\:[M^2(L,T,p)] & = & E\: \left[ E\: [ M^2(L,T,p) \; | \; X_1 ] \right] \nonumber \\
 & = & E\: [M^2(L,T,p) \; | \; X_1 =1] (1-p) \; \nonumber \\ 
  & & \ \ \ \ \ \ \ \ \ \ + \: E[M^2(L,T,p) \; | \; X_1 \neq 1] p \nonumber \\
& = & (1-p) E\:[M^2(L-1, T,p)] \: + \: p E( 1 \: + \: M(L-T, T,p) )^2 \nonumber \\
& = & (1-p) E\:[M^2(L-1, T,p)] \: + \nonumber \\ 
  & &  \ \ \ \ \ \ \ \ \ \  \: p E( 1 \: + \: M^2(L-T, T,p) \: + \: 2 M(L-T, T,p)) \nonumber \\ 
& & 
\end{eqnarray}
Thus we get
\begin{equation}
G(L,T,p) = (1-p) G(L-1, T,p) \: + \: p(1 \: + \: G(L-T, T,p) \: + \: 2 F(L-T, T,p))
\label{eq:var-rec1}
\end{equation}

Solving the above, we get the second moment of $M$. Let, $V(L,T,p)$ be the 
variance of $M(L,T, p)$. Then we have 
\begin{equation}
V(L,T,p) = G(L,T,p) \: - \: (F(L,T,p))^2 
\label{eq:var}
\end{equation}

Once we have the mean and variance we can bound the probability that the number 
of non-overlapped occurrences is beyond something. For example, we can use 
Chebyshev inequality as
\begin{equation}
\mbox{Pr}\left[|M(L.T,p) - F(L,T,p)| > k \sqrt{V(L,T,p)}\right] \leq \frac{1}{k^2}
\label{eq:cheb}
\end{equation}
for any positive $k$. \footnote{This may be a loose bound. We may get better bounds by 
using central limit theorem based arguments. But for our purposes here, this 
is not very important. Also, as we shall see from the empirical results 
presented in the next section, this bound seems to be adequate}. 
Such bounds can be used for test of statistical significance as explained below. 

\subsection{ Test for statistical significance} 

Suppose we are considering $n$-node episodes. Let the allowable Type I error for 
the test be $\epsilon$. Then what we need is a threshold, say, $m_{th}$ for 
which we have 
\begin{equation}
\mbox{Pr}_n\left[f_{epi} \geq  m_{th} \right] \leq \epsilon,  
\label{eq:test1}
\end{equation}
where $f_{epi}$ is the frequency of any $n$-node episode and 
$\mbox{Pr}_n$ denotes  probability under the null hypothesis models. 

This would imply that if we find a $n$-node episode with frequency greater than 
$m_{th}$ then, with $(1 - \epsilon)$ confidence we can reject our null hypothesis 
and hence assert that the discovered episode represents `strong' interactions    
among those neurons. 

Now the above can be used for assessing statistical significance of any 
episode as follows. Suppose 
we are considering an $n$-node (serial) episode. Let the first node of this episode 
have event type $A$. (That is, it corresponds to neuron $A$). Let $\rho_A$ be the probability 
that $A$ will spike in any interval of length $\Delta T$. (We will fix $\Delta T$ by 
the time resolution being considered). 
Let $\epsilon$ be the prescribed confidence level. Let $k$ be such that 
$k^2 \geq \frac{1}{\epsilon}$. Fix $p= \rho_A (e_0)^{n-1}$. Let $T$ be the sum of all inter-event 
delay times in the episode. Let $L$ be the total length of data (as time span in units 
of $\Delta T$).  

Our null hypothesis is that the conditional probability for any pair of neurons 
is less than $e_0$. Further, our random variable $M$ is such that its probability 
of taking higher values increases monotonically with $p$. 
Hence, with the above $p$, the probability of $M(L,T,p)$ being greater than any value 
is an upper bound on the probability of the episode frequency being greater than 
that value under any of the models in our null hypothesis. 

Thus, a threshold  for significance 
is $m_{th} = F(L,T,p) + k \sqrt{V(L,T,p)}$ because, from 
eq.~(\ref{eq:cheb}) we have 
\begin{equation}
\mbox{Pr}\left[M(L.T,p) \geq  F(L,T,p) + k \sqrt{V(L,T,p)}\right] \leq \frac{1}{k^2} \leq 
\epsilon. 
\label{eq:cheb1}
\end{equation}

Though we do not have closed form expressions for $F$ and $V$, using our recurrence 
relations, we can calculate $F(L,T,p)$ and $V(L,T,p)$ for any given values of 
$L,T,p$ and hence can calculate the above threshold. 
The only thing unspecified for this calculation is how do we get $\rho_A$. 
We can obtain $\rho_A$ by either estimating 
the average rate of firing for this neuron from the data or from other prior knowledge. 

Thus, we can use 
eq.~(\ref{eq:cheb1}) either for assessing the significance of a specific 
$n$-node episode or for fixing a threshold of any $n$-node episode in our datamining 
algorithm. In either case, this allows us to deduce the `strong connections' (if any) 
in the neural system being analyzed by using our datamining method. 

We can summarize the the test of significance as follows. Suppose the allowed type-I error 
is $\epsilon$. We choose integer $k$ such that $\epsilon < \frac{1}{k^2}$. Suppose we want 
to assess the significance of a n-node sequential pattern with the total delay being $T$ 
based on its count. Suppose $e_0$ is the bound we 
use in our null hypothesis. Let $L$ be the total data length in time units. Let $\rho$ be 
the average firing rate of the first neuron in the data. Let $p=\rho (e_0)^{n-1}$. We calculate 
$F(L,T,p)$ and $V(L,T,p)$ using (\ref{eq:rec1}), (\ref{eq:var-rec1}) and (\ref{eq:var}). Then 
the pattern is declared significant if its count 
 exceeds $F(L,T,p) + k \sqrt{V(L,T,p)}$.\footnote{This threshold for a pattern to be significant 
depends on the size of the pattern with smaller size patterns needing higher count to be 
significant, as is to be expected. This also adds to the efficiency of our data mining algorithm for 
discovering sequential patterns. In the level-wise procedure described earlier, we would have higher 
thresholds for smaller size patterns thus further mitigating the combinatorial explosion in 
the process of frequent episode discovery.}

{\em We like to emphasize that the threshold frequency (count)  given above for an episode 
to be significant (and hence represent strong interactions) is likely to be larger 
than that needed. This is because it is obtained through a Chebyshev bound which 
is often loose.}
Thus, for example, if we choose $e_0=0.4$ then some strong connections which 
may result in the effective conditional probability value of up to 0.5  may 
not satisfy the test of significance at a particular significance level. This, in 
general, is usual in any hypothesis testing framework. In practice, we found that 
we can very accurately discover all connections whose strengths in terms of the 
conditional probabilities are about 0.2 more than $e_0$ at 5\% confidence level. 
At $\epsilon = 0.05$, the threshold is about 4.5 standard deviations above the 
mean. In a specific application, for example, if we feel that three standard 
deviations above the mean is a good enough threshold, then correspondingly we will 
be able to discover even those connections whose effective conditional probability 
is only a little above $e_0$. 

This test of significance allows us to rank order the discovered patterns. For this, 
we run our datamining method with 
different thresholds corresponding to different $e_0$ values. Then, by looking at the 
sets of episodes found at different $e_0$ values, we can essentially rank order 
the strengths of different connections in the underlying system. Since any manner of 
assigning numerical values to strengths of connections is bound to be somewhat arbitrary, 
this method of rank ordering different connections in terms of strengths can be much 
more useful in analyzing microcircuits.  

We illustrate all these through our simulation experiments in section~\ref{sec:simu}. 

\subsection{Extension to the model}

So far in this section we have assumed that the individual  delays and hence 
the span of an episode, $T$, to be constant. In practice, even if delay is 
random and varies over a small interval around $T$, the threshold we calculated 
earlier would be adequate. In addition to this, it is possible to  
 extend our model to take care of some random variations 
in such delays. 

Since we have assumed that $\Delta T$ is the time resolution at which we are working, 
it is reasonable to assume that the  delay $T$ is actually specified in units of 
$\Delta T$. Then we can think of the delay as a random variable taking 
values in a set $\{ T-J, \; T-J+1, \; \cdots, T+J \}$ where $J$ is a small (relative to T) 
integer. For example, suppose the delay is uniformly   
distributed over $\{ T-1, \; T, \; T+1 \}$. 
 Now we can change our model as follows:

 The $\{X_i, \ i=1,2,\ldots \}$ will now be {\em iid} random 
variables with distribution 
\begin{eqnarray}
\mbox{Prob}[X_i = 1]  & = &  1 - p \nonumber \\
\mbox{Prob}[X_i = T-1]  =    
\mbox{Prob}[X_i = T]  = 
\mbox{Prob}[X_i = T+1]  &=&  \frac{p}{3} \nonumber 
\label{eq:xi-new}
\end{eqnarray}
where we now assume that $T > 2$. 

We will define $N$ as earlier by eq.~(\ref{eq:N}). We will now define $Z$ as the number 
of $X_i$ out of first $N$ that {\bf do not} take value 1. In terms of this $Z$, we 
will define $M$ as earlier by eq.~(\ref{eq:M}). 

Now it is easy to see that our $M(L,T,p)$ would again be the random variable 
corresponding to number of non-overlapped occurrences in this new scenario where 
there are random variations in the  delays. Now the recurrence relation for 
$F(L,T,p)$ would become   
\begin{eqnarray}
F(L,T,p) &=& (1-p) F(L-1, T, p) + \nonumber \\
 & & \hspace*{-1.5cm} p \left( 1 + \frac{1}{3}(F(L-T+1, L, p)+F(L-T, L, p) + F(L-T-1, L, p))\right) \nonumber \\
& & 
\label{eq:rec1-ex}
\end{eqnarray}

The recurrence relation for variance of $M(L,T,p)$ can also be similarly derived. Now, we can 
easily implement the significance test as derived earlier. While the recurrence relations 
are a little more complicated, it makes no difference to our method of significance 
analysis because these recurrence relations are anyway to be solved numerically. 

It is easy to see that this method can, in principle, take care of any distribution of the 
total  delay (viewed as a random variable taking values in a finite set) by 
modifying the recurrence relation suitably.

\section{Simulation Experiments}
\label{sec:simu}

In this section we describe some simulation experiments to show the effectiveness of 
our method of statistical significance analysis. We show that our stochastic model 
properly captures our counting process and that the frequency threshold we calculate 
is effective for separating connections that are `strong' (in the sense of  
conditional probabilities). We also show 
that our frequency can properly rank order the strengths of connections 
in terms of conditional probabilities. As a matter of fact, our results provide good 
justification for saying that conditional probabilities provide a very good scale 
for denoting connection strengths. For all our experiments we choose synthetically 
generated spike trains. This is  because then we know the ground truth about 
connection strengths and hence can test the validity of our statistical theory. 
For the simulations we use a data generation scheme 
where we model the spiking of each neuron as an inhomogeneous Poisson process on 
which is imposed an additional constraint of refractory period. (Thus the actual 
spike trains are not truly Poisson even if we keep the rate fixed). 
The inhomogeneity in the Poisson process are due to
 the instantaneous firing rates being modified based on total 
input spikes received by a neuron through its synapses. 

We have shown elsewhere \cite{PSU2008,archive-report} that our datamining algorithms 
are very efficient in discovering interesting patterns of firings from spike 
trains and that we can discover patterns of more than ten neurons also. Since in this 
paper the focus is on statistical significance of the discovered patterns, we would 
not be presenting any results for showing the computational efficiency of the method. 

\subsection{Spike data generation}

We use a simulator for generating the spike data from a network of interconnected 
neurons. Let $N$ denote the number of neurons in the network. The spiking of each 
neuron is modelled as an inhomogeneous Poisson process whose rate of firing is 
updated at time intervals of $\Delta T$. (We normally take $\Delta T$ to be 1ms). 
The neurons are interconnected by synapses and each synapse is characterized by 
a delay (which is in integral multiples of $\Delta T$)  and a weight which is a 
real number. All neurons also have a refractory period. The rate of the Poisson 
process is varied with time as follows.
\begin{equation}
\lambda_j(k) = \frac{K_j}{1 + \exp{(-I_j(k) + d_j)}}
\label{eq:lambda-update}
\end{equation}
where $\lambda_j(k)$ is the firing rate of $j^{th}$ neuron at time $k \Delta T$, 
 and $K_j, d_j$ are two parameters.  
 $I_j(k)$ is the total input into $j^{th}$ neuron at time $k \Delta T$ and it is 
given by
\begin{equation} 
I_j(k) = \sum_i O_i(k) w_{ij}
\label{eq:input}
\end{equation}
where $O_i(k)$ is the output of $i^{th}$ neuron  (as seen by
the $j^{th}$ neuron) at time $k \Delta T$
 and $w_{ij}$ is the weight of synapse from $i^{th}$ to $j^{th}$ neuron.
$O_i(k)$ is taken to be the number of spikes by the $i^{th}$ neuron in the time
interval $(\;(k-h_{ij}-1) \Delta T, \ (k-h_{ij}) \Delta T]$ where $h_{ij}$ represents the
 delay (in units of $\Delta T$) for the synapse from $i$ to $j$.
The parameter $K_j$ is chosen based on the dynamic range of firing rates that we 
need to span. The parameter $d_j$ determines the `background' spiking
rate, say,  $\lambda_{0j}$. This is the firing rate of the $j^{th}$ neuron 
under zero input. After choosing a suitable value for  $K_j$, 
we fix the value of $d_j$ based on this 
background firing rate specified for each neuron. 

We first build a network that has many random interconnections with low weight values 
and a few strong interconnections with large weight values. We then generate spike 
data from the network and show how our method can detect all strong connections. 
To build the network we specify the background firing rate (which we normally keep 
same for all neurons) which then fixes the value of $d_j$ in (\ref{eq:lambda-update}). 
We specify all weights in terms of conditional probabilities. Given a conditional 
probability we first calculate the needed instantaneous firing rate so that probability 
of at least one spike in the $\Delta T$ interval is equal to the specified 
conditional probability. Then, using (\ref{eq:lambda-update}) and 
(\ref{eq:input}), 
we calculate the value of $w_{ij}$ needed so that the receiving neuron ($j$)  
reaches this instantaneous rate given that the sending neuron ($i$) spikes once 
in the appropriate interval and assuming that input into the receiving neurons from 
all other neurons is zero. 

We note here that the background firing rate as well as the effective conditional 
probabilities in our system would have some small random variations. As said above, 
we fix $d_j$ so that on zero input the neuron would have the background firing rate. 
However, all neurons would have synapses with randomly selected other neurons and 
the weights of these synapses are also random. Hence, even in the absence of any 
strong connections, the firing rates of different neurons keep fluctuating around the 
background rate that is specified. Since we choose random weights from a zero mean 
distribution, in an expected sense we can assume the input into a neuron to be 
zero and hence the average rate of spiking would be the background rate specified. 
We also note that the way we calculate the effective weight for a given conditional 
probability is also approximate and we chose it for simplicity. If we specify 
a conditional probability for the connection from $A$ to $B$, then, the method stated 
in the previous paragraph fixes the weight of connection so that the probability of 
$B$ firing at least once in an appropriate interval given that $A$ has fired is equal 
to this conditional probability {\em when all other input into $B$ is zero}. But since 
$B$ would be getting small random input from other neurons also, the effective 
conditional probability would also be fluctuating around the nominal value specified. 
Further,  even if the random weights have zero mean, the fluctuations in the 
conditional probability may not have zero mean due to the nonlinear sigmoidal 
relationship in (\ref{eq:lambda-update}). The nominal conditional probability 
value determines where we operate on this sigmoid curve and  that determines 
the bias in the excursions in conditional probability for equal fluctuations in either 
directions in the random input into the neurons. We consider this as a noise in the 
system and show that our method of significance analysis is still effective.

The simulator is run as follows. First, for any neuron we fix a fraction (e.g., 25\%) of 
all other neurons that it is connected to. The actual neurons that are connected to any 
neuron are then selected at random using a uniform distribution. We fix the  delays 
and background firing rates for all neurons. 
We then assign random weights to 
connections by choosing uniformly from an interval. In our simulation experiments we 
specify this range in terms of conditional probabilities. For example suppose the 
background firing rate is 20 Hz. Then with $\Delta T = 1ms$, the probability of 
firing in any interval of length $\Delta T$ is (approximately) 0.02. Hence a conditional 
probability of 0.02 would correspond to a weight value of zero. Then a range of 
conditional probabilities such as $[0.01, \ 0.04]$ (increase or decrease by a 
factor of 2 in either direction) would correspond to a weight range around zero. 
After fixing these random weights, we incorporate a few strong connections which 
vary in different simulation experiments. These weight values are also specified in 
terms of conditional probabilities. We then generate a spike train by simulating all 
the inhomogeneous Poisson processes where rates are updated every $\Delta T$ time instants. 
We also fix refractory period for neurons (which is same for all neurons).
Once a neuron is fired, we will not let it fire till the refractory 
period is over.

\subsection{Results}

For the results reported here we used a network of 100 neurons with the nominal firing rate being 
20 Hz. Each neuron is connected to 25 randomly selected neurons with the effective conditional probability 
of the connection strength ranging over $[0.01, \ 0.04]$. With 20Hz firing rate and 1ms time resolution, the 
effective conditional probability when two neurons are independent is 0.02. Thus the random connections 
have conditional probabilities that vary by a factor of two on either side as compared to the independent case. 
We then incorporated some strong connections among some neurons. For this we put in one 3-node episode, three 
4-node episodes, three 5-node episodes and one 6-node episode with different strengths for the connections. 
The connection strengths are so chosen so that we have enough number of 3-node and 4-node episodes (as 
possibly subepisodes of the embedded episodes) spanning 
the range of conditional probabilities from 0.1 to 0.8. All synaptic connections have a delay of 5ms. 
Using our simulator described earlier, we generated 
spike trains for 20 sec of time duration (during which there are about 50,000 spikes typically), 
 and obtained the counts of non-overlapped occurrences  of episodes 
of all sizes using our datamining algorithms. In all results presented below, all statistics are calculated 
using 1000 repetitions of this simulation. Typically, on a data sequence for 20 Sec duration, the mining 
algorithms (run on a dual-core Pentium machine) take about a couple of minutes. 

\begin{figure}
\centering
\includegraphics[scale=0.75,clip]{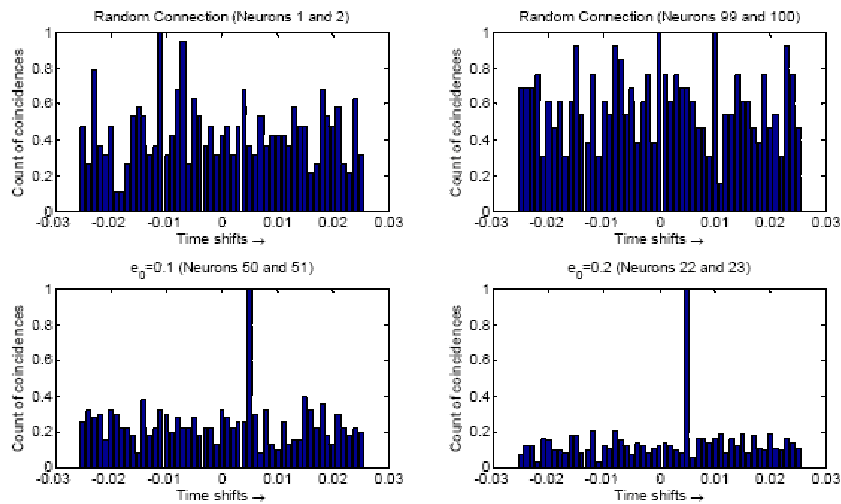}
\caption{Normalized cross correlograms (obtained through 1000 replications) for four different pairs of 
neurons. The top two panels show pairs of neurons with weak interconnections while the two bottom panels 
show neuron pairs with strong interconnections. For neurons pairs in the bottom two panels, the 
cross correlogram shows strong peak.} 
\label{fig:corr}
\end{figure}

As explained earlier, in our simulator, the rate of the Poisson process (representing the spiking of a 
neuron) is updated every 1ms based on the actual spike inputs received by that neuron. This would, in general, 
imply that many pairs of neurons (especially those with strong connections) are not spiking as independent 
processes. Fig.~\ref{fig:corr} shows this for a few pairs of neurons. The figure shows the cross correlograms 
(with bin size of 1 ms and obtained using 1000 replications) for pairs of neurons that have 
 weak connections and for pairs of neurons that have 
strong connections. There is a marked peakiness in the cross correlogram for neurons with strong interconnections, 
as expected.   

\begin{figure}
\centering 
\begin{tabular}{cc}
\includegraphics[scale=0.5,clip]{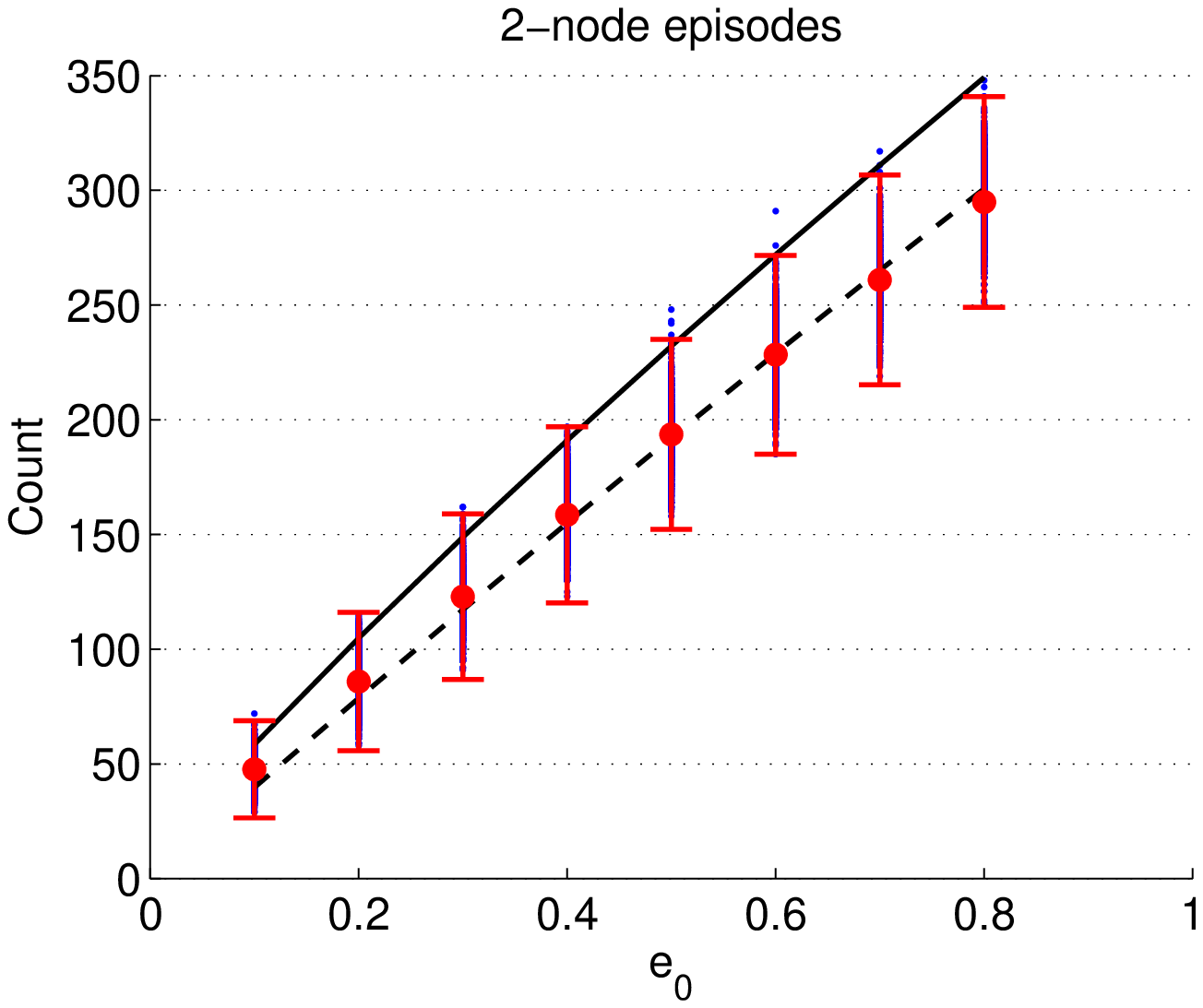} &
\includegraphics[scale=0.5,clip]{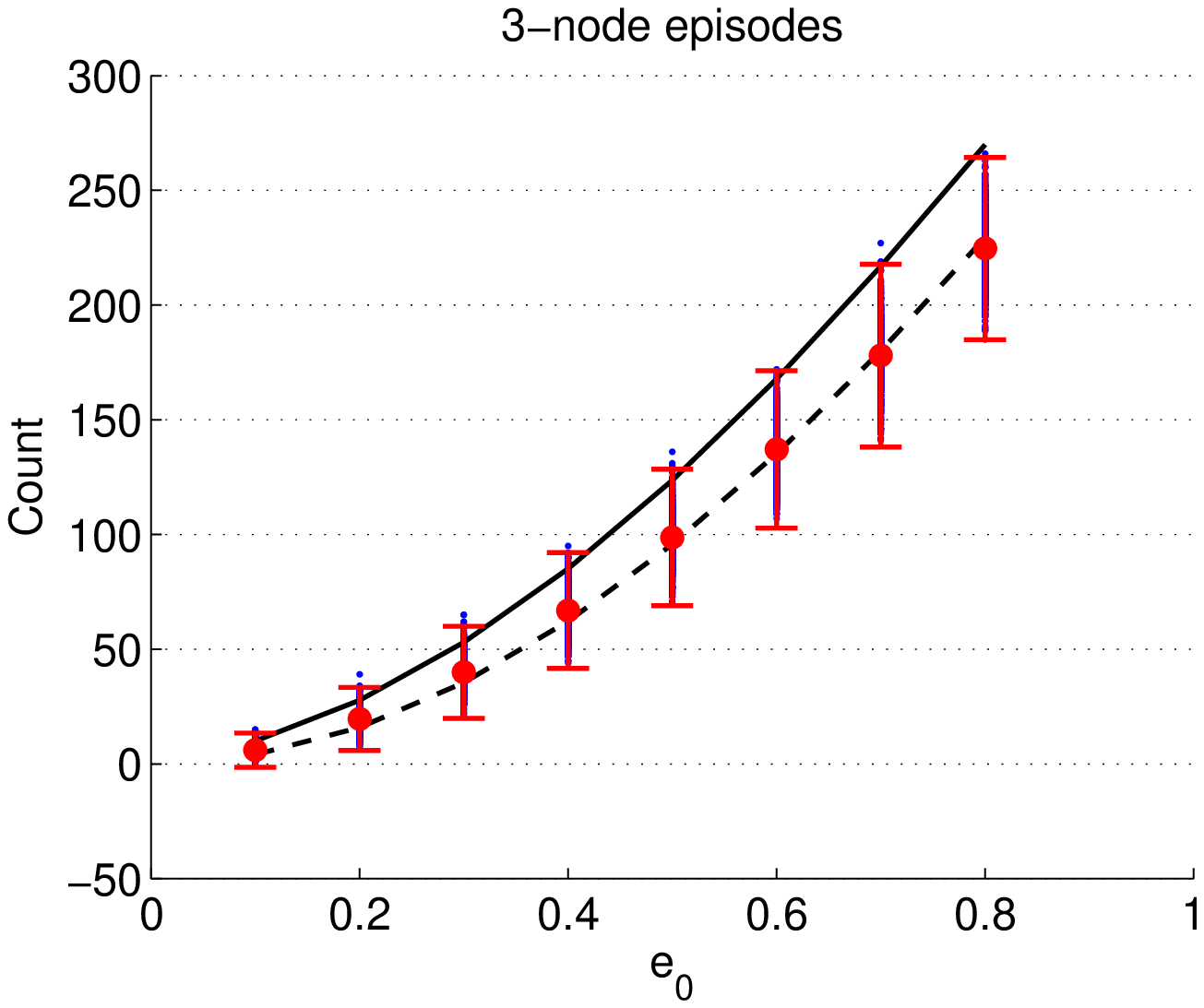} \\
\includegraphics[scale=0.5,clip]{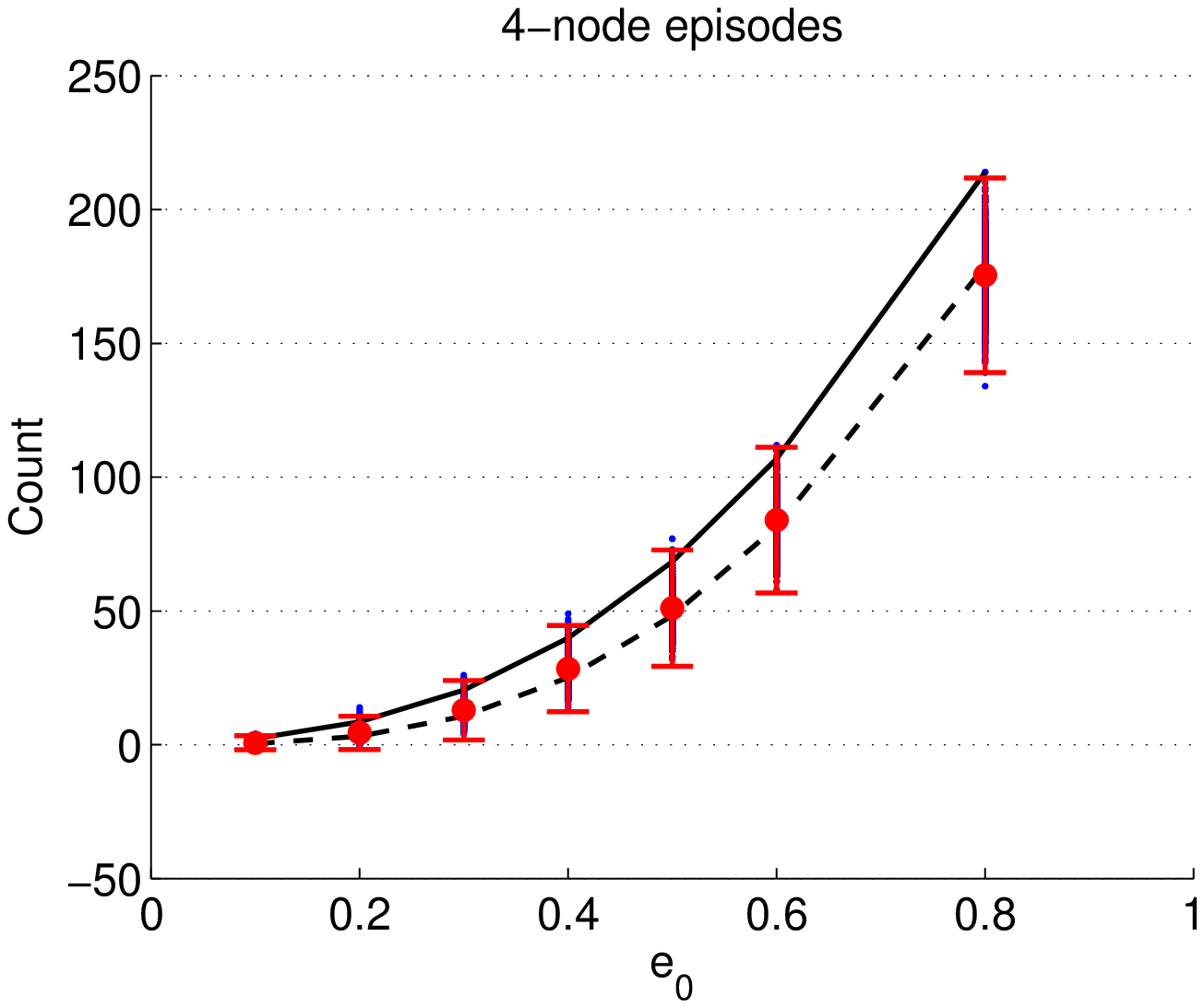} &
\includegraphics[scale=0.5,clip]{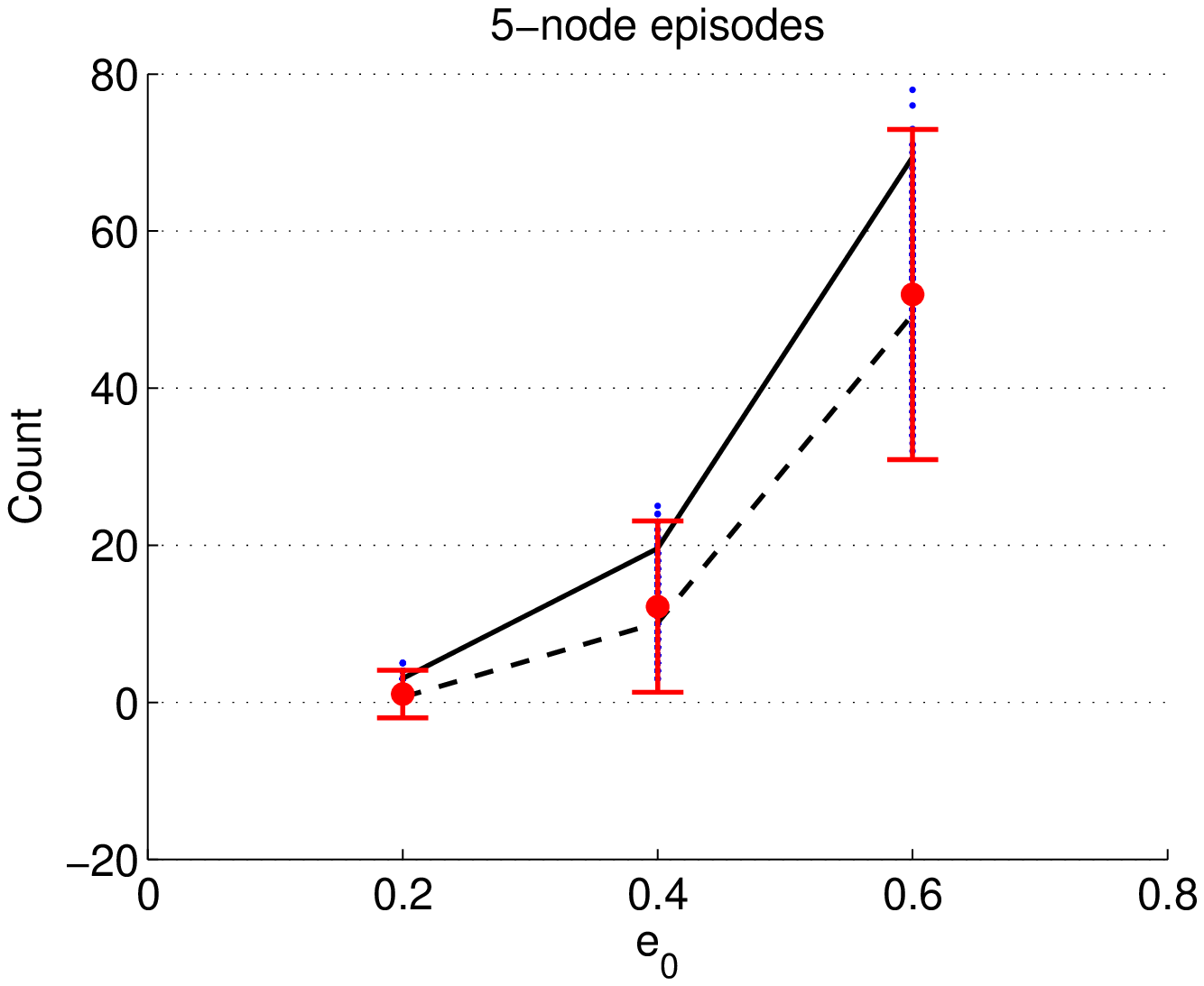} \\
\end{tabular}
\caption{The analytically calculated values for the mean (i.e., $F$)  
 and the mean plus 3 $\sigma$ (i.e., $F+3\sqrt{V}$),  
as a function of the connection strength in terms of conditional probabilities. The top two panels 
show plots for 2-node and 3-node patterns and bottom panels show plots for 4-node and 5-node patterns. For each 
value of the conditional probability, the actual counts as obtained by the algorithm are also shown 
 These are obtained through 1000 replications. For these experimental counts, the mean value as well as the 
$\pm 3 \sigma$ range (where $\sigma$ is the data standard deviation) are also indicated. 
As can be seen, the calculated 
value of $F$ well captures the mean of the non-overlapped counts. The 
$F+3\sqrt{V}$  line captures most of the count distribution.   }
\label{fig:acc-3-4-5}
\end{figure}

Fig.~\ref{fig:acc-3-4-5} shows that our theoretical model for calculating the mean and variance of 
of the non-overlapped count (given by $F$ and $V$ determined through eqns. (\ref{eq:rec1}) and 
(\ref{eq:var}) ) are accurate. 
The figure shows  plot of the mean ($F$) and 
mean plus three times standard deviation ($F+3\sqrt{V}$) for different values of the connection 
strength in terms of conditional probabilities ($e_0$), for the different episode sizes. Also shown are  
 the actual counts obtained for episodes of that size with different $e_0$ values. As is 
easily seen, the theoretically calculated mean and standard deviations are very accurate. 
Notice that 
most of the observed counts are below the $F+k\sqrt{V}$ threshold for $k=3$ even though this  
corresponds to a Type-I error of just over 10\%. Thus our statistical test with 
$k=3$ or $k=4$ should be quite effective. 


\begin{figure}
\centering
\includegraphics[scale=0.5]{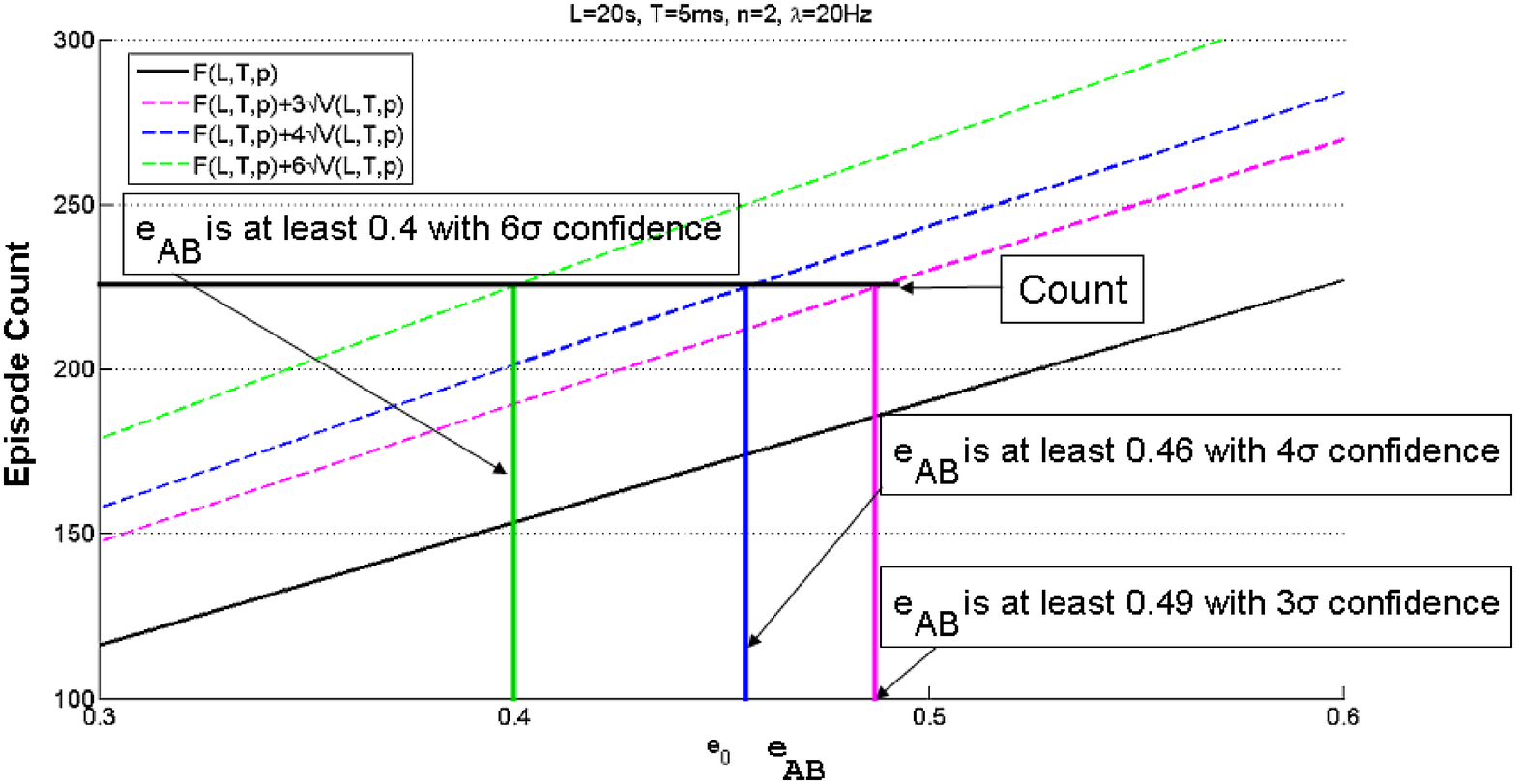}
\caption{Illustration of inferring of a connection strength based on observed count for a 
pattern. Given the curves of mean $F$, and the various levels of threshold ($F+3\surd{V}$, $F+4\surd{V}$, and $F+6\surd{V}$), we can `invert' the 
observed count to obtain a connection strength at which the observed count makes the episode just significant at a particular level. 
We call this the inferred connection strength based on the observed count. }
\label{fig:infer-strength}
\end{figure}

As explained earlier, using the formulation of our significance test we can infer 
 a (bound on the) connection strength in terms of conditional probability  
based on the observed count. For this, given observed count of a sequential  
pattern or episode, we ask what is the value of the strength or  
conditional probability of the connection at which this count is the threshold 
as per our significance test. This is illustrated in Fig.~\ref{fig:infer-strength}. 
For an $n$-node episode if the inferred strength is $q$ then we can assert (with the appropriate 
confidence) that it is highly unlikely for this episode to have this count if connection strength between 
every pair of neurons is less than $q$ 

\begin{figure}
\centering 
\begin{tabular}{cc}
\includegraphics[scale=0.5]{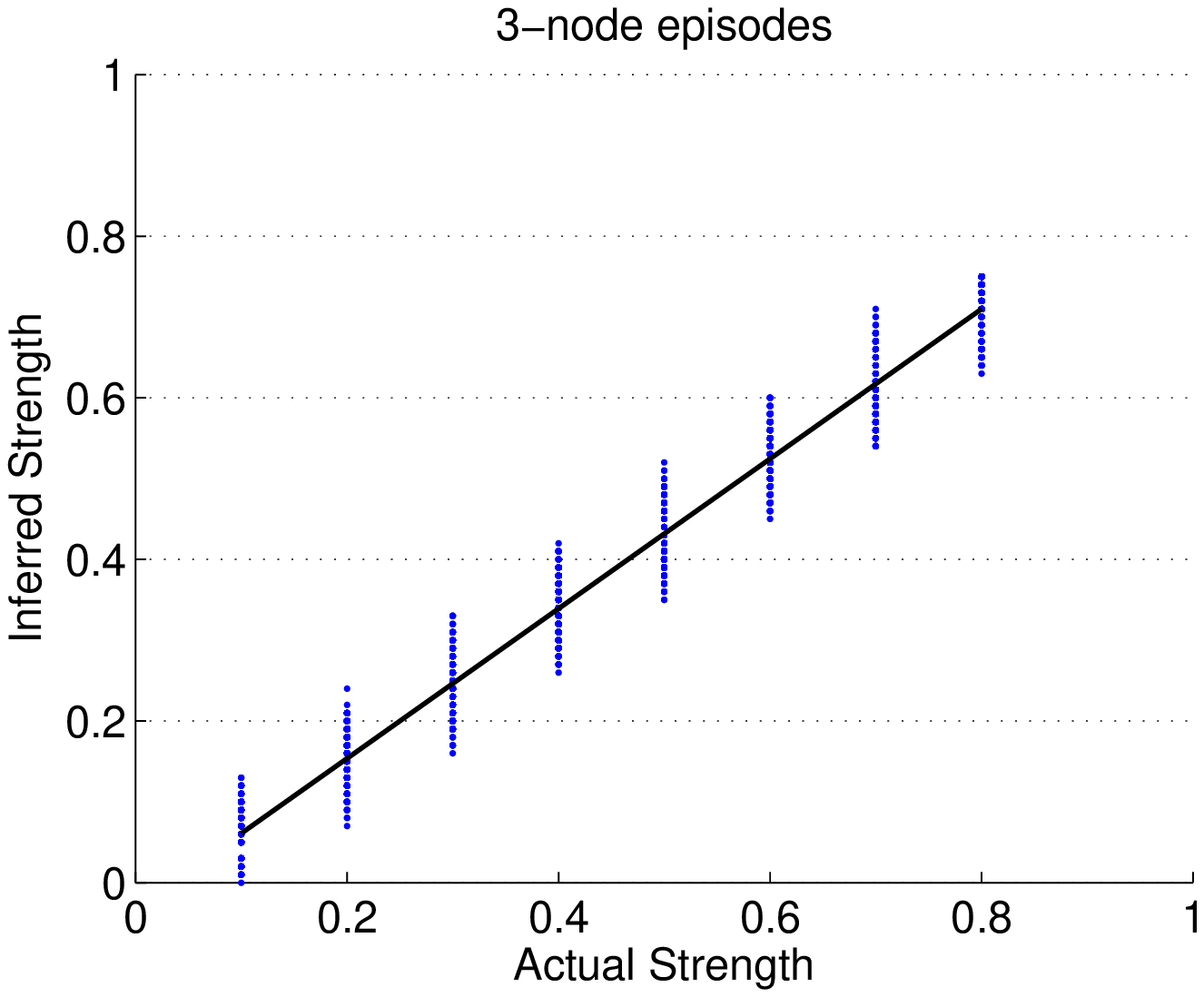} &
\includegraphics[scale=0.5]{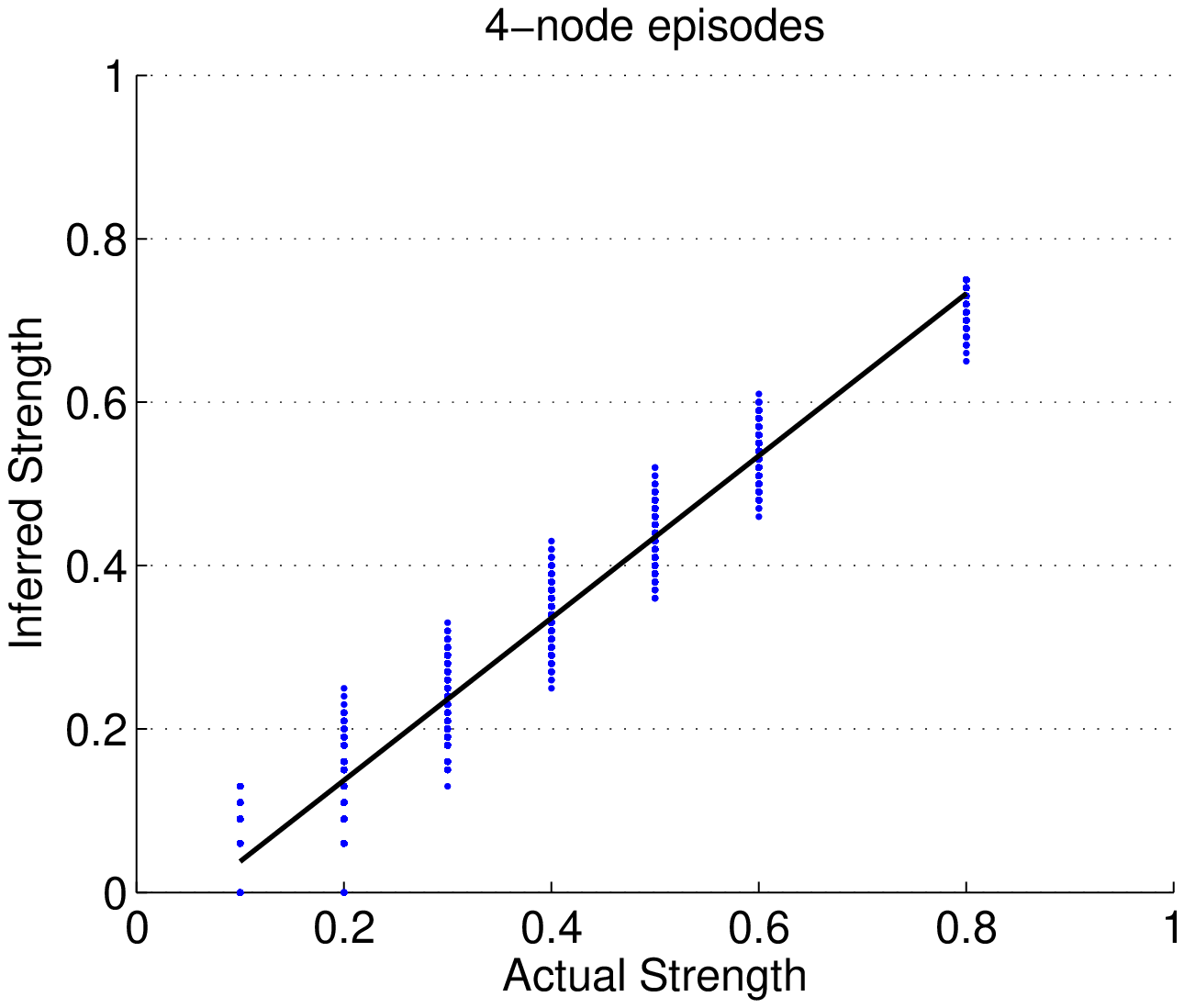} \\
\end{tabular}
\caption{Plot of the actual value of the conditional probability used in simulation versus the  
value inferred from our test of significance as explained in text. (See fig.~\ref{fig:infer-strength}. 
For each value we do 1000 replications and the 
different inferred values are shown as a point cloud. Also shown is a best fit line. The two panels show results 
for episodes of size 3 and size 4. Our method is quite effective in inferring a connection strength 
based on our count. }
\label{fig:good-infer-strength}
\end{figure}

In Fig.~\ref{fig:good-infer-strength} we show how good is this mechanism for inferring the strength of connection. 
Here we plot the actual value of the strength of connection in terms of the conditional probability as used in the 
simulation against the inferred value of this strength from our theory based on the actual observed value of count. 
For each value of the conditional probability, we have 1000 replications and these various inferred values are shown as 
point clouds. Since the theory is based on a bound, the inferred value would always be lower than the actual strength.
However, the results in this figure show the effectiveness of our approach to determining significance of 
sequential patterns based on counting the non-overlapped occurrences.   We emphasize here that this 
inferred value of strength is based on our significance test and 
there is no  estimation of any conditional probabilities.

\begin{figure}
\centering
\begin{tabular}{cc}
\includegraphics[scale=0.5]{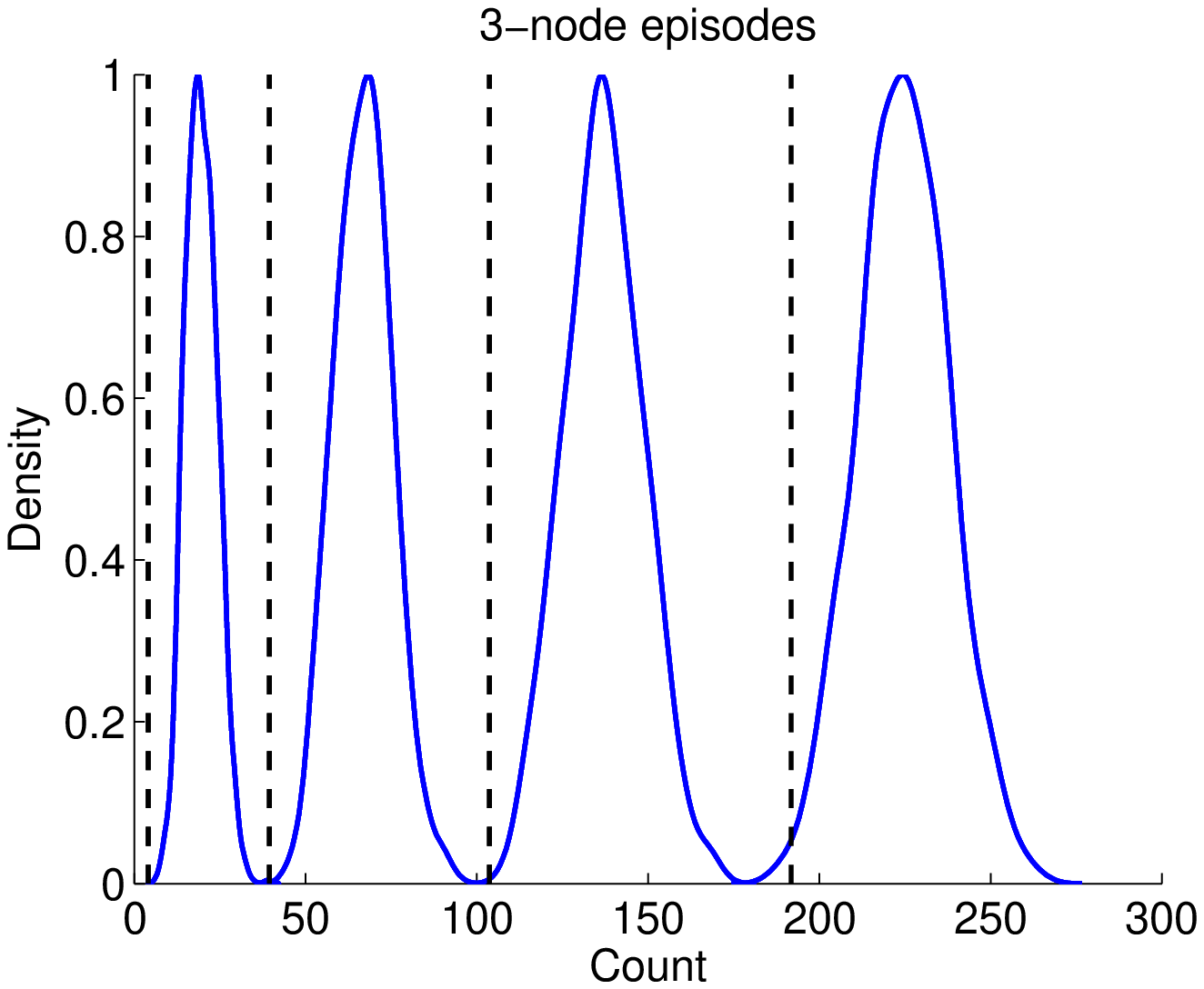} & 
\includegraphics[scale=0.5]{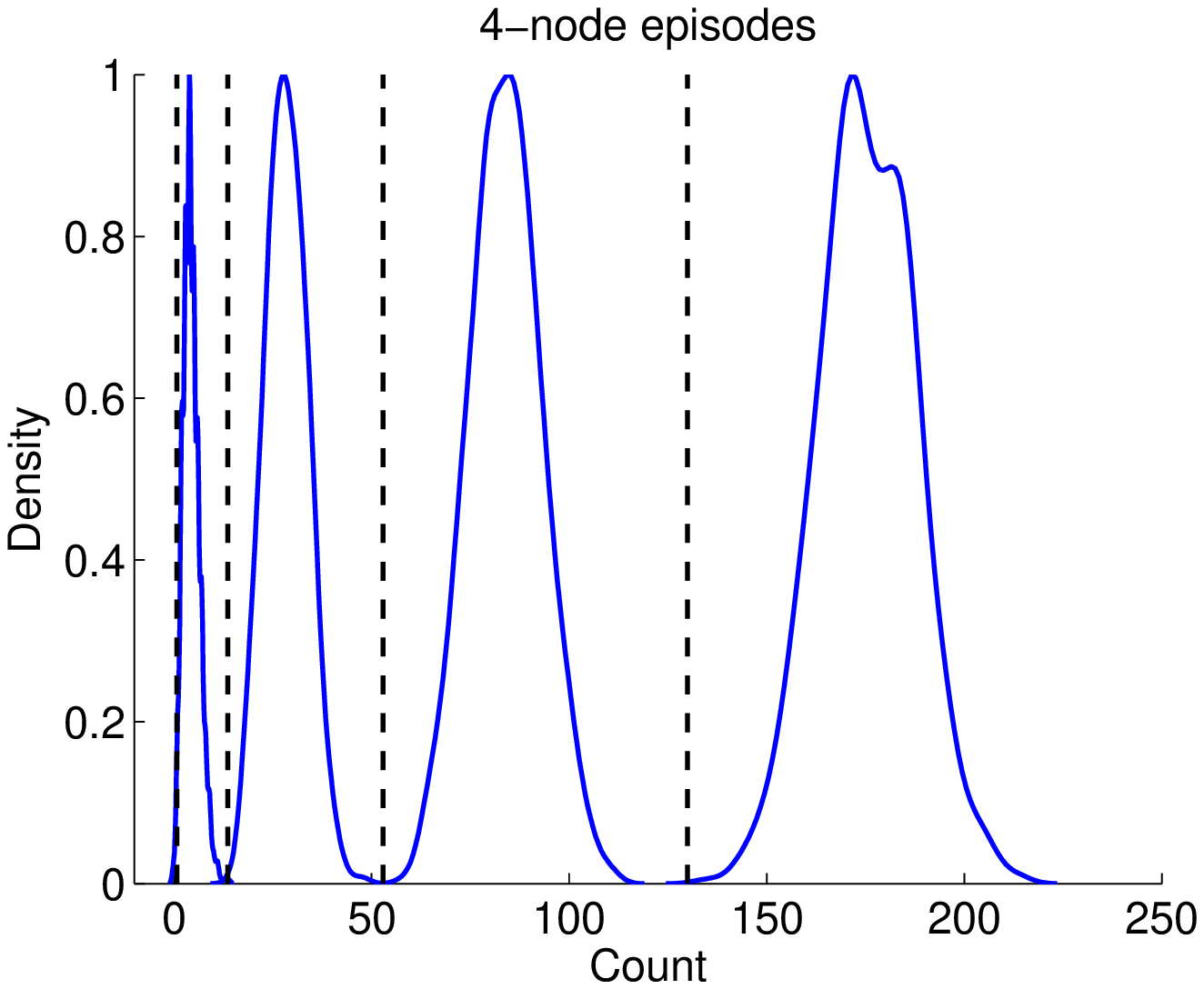} \\
\includegraphics[scale=0.5]{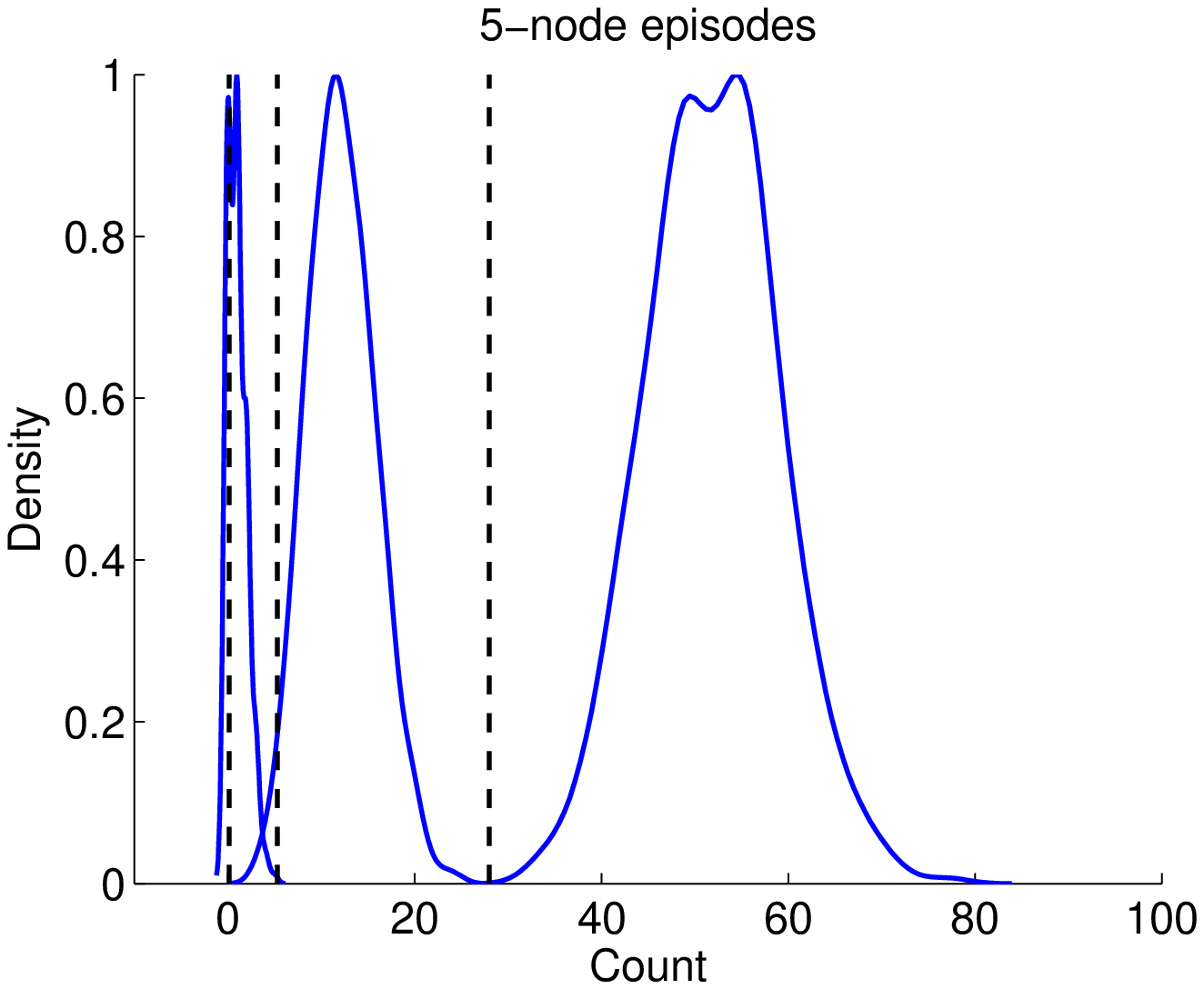} & \\
\end{tabular}
\caption{Plot showing the ability of our method of statistical significance test at inferring relative strengths 
of different patterns. Top two panel shows the distribution of counts (over 1000 replications) for four 3-node 
and 4-node  
episodes with connection strengths corresponding to 0.2, 0.4, 0.6 and 0.8. The dashed lines are the thresholds 
on counts under our significance test (with $k=3$) corresponding to $e_0$ values of 0.05, 0.25, 0.45 and 0.65.
The bottom panel shows distributions for 5-node episodes with strengths 0.1, 0.3 and 0.5 with thresholds 
corresponding to $e_0$=0.05, 0.15 and 0.35.  
 Since our test is based on Chebyshev inequality, it overestimates the needed count. However, it is easy to 
see that we can detect significant episodes corresponding to different strengths by varying the $e_0$ in our 
null hypothesis. 
As can be seen, our method is able to reliably 
infer the relative strengths of different patterns.  }
\label{fig:rankorder-3-4-5}
\end{figure}

Finally, we present some results to illustrate the ability of our significance test to correctly rank order 
different sequential patterns or episodes that are significant. For this we show the distribution of 
counts for sequential patterns or episodes of different strength along with the thresholds as calculated 
by our significance test when the value of $e_0$ in the null hypothesis is varied. These results are shown for 
3-node, 4-node  and 5-node episodes  in fig.~\ref{fig:rankorder-3-4-5}.  
 From the figure we can see that, by choosing a particular $e_0$ value in the 
null hypothesis, our test will flag only episodes corresponding to strength higher than $e_0$ as significant. Thus, 
by varying $e_0$ we can rank-order different significant patterns that are found by the mining algorithm. 
We note here that our threshold actually overestimates the count needed because it is based on a loose bound. 
However, these results show that we can reliably infer the relative strengths of different sequential patterns. 

\section{Discussion}
\label{sec:dis}

In this paper we addressed the problem of detecting statistically significant 
sequential patterns in multi-neuronal spike train data. We employed an efficient 
datamining algorithm that detects all frequently occurring sequential patterns 
with prescribed inter-neuron delays. A pattern is frequent if the number of 
non-overlapping occurrences of the pattern is above a threshold. The strategy 
of counting only the non-overlapped occurrences rather than all occurrences makes 
the method computationally attractive. The main contribution of the paper is a 
new statistical significance test to determine when the count obtained by our 
algorithm is statistically significant. Or, equivalently, the method gives a 
threshold for different patterns so that the algorithm can detect only the 
significant patterns. 

The novelty in assessing the significance in our approach is in the structure of 
the null hypothesis. The idea is to use conditional probability as a mechanism to 
capture strength of influence of one neuron on another. Our null hypothesis is specified in terms 
of a (user-chosen) bound on the conditional probability that $B$ will fire after 
a specified delay given that $A$ has fired, for any pair of neurons $A$ and $B$. Thus this compound null 
hypothesis includes many models of inter-dependent neurons where the influences 
among neurons are `weak' in the sense that all such pairwise conditional probabilities 
are below the bound. Being able to reject such a null hypothesis makes a stronger 
case for concluding that the detected patterns represent significant functional 
connectivity. Equally interestingly, such a null hypothesis allows us to rank order 
the different patterns in terms of their strengths of influence. If we chose this 
bound $e_0$ to be the value of the conditional probability when the different neurons 
are independent, then we get the usual null hypothesis of independent neuron model. 
But since we can choose the $e_0$ to be much higher, we can decide which patterns are 
significant at different levels of $e_0$ and hence get an idea of the strength of 
interaction they represent. Thus, the method presented here extends the current 
techniques of significance analysis. 

While we specify our null hypothesis in terms of a bound on the conditional 
probability, note that we are not in any way estimating such conditional 
probabilities. Estimating all relevant conditional probabilities would be 
computationally intensive. Since our algorithm counts only non-overlapped 
occurrences and also uses the datamining idea of counting frequencies for only 
the relevant candidate patterns, our counts do not give us all the pair-wise conditional 
probabilities. However, the statistical analysis presented here allows us to 
obtain thresholds on the non-overlapped occurrences possible (at the 
given confidence level) if all the conditional probabilities are below our bound. 
This is what gives us the test of significance. 

We presented a method for bounding the probability that, under the null hypothesis, 
a pattern would have more than some number of non-overlapped occurrences. Because 
we are counting non-overlapped occurrences, we are able to capture our counting 
process in an interesting model specified in terms of sums of independent random variables. 
This model allowed us to get recurrence relations for mean and variance of the 
random variable representing our count under the null hypothesis which allowed us 
to get the required threshold using Chebyshev inequality. While this may be a loose 
bound, as shown through our simulation results, the bound we calculate is very 
effective. 

Our method of analysis is quite general and it can be used in 
situations other than what we considered here. By choosing the value of $p$ 
in eq.(\ref{eq:xi}) appropriately we can realize this generality in the model.

As an illustration of this we will briefly describe one extension of the model. 
In the method presented, while analyzing significance of a pattern 
$A\stackrel{T_1}{\rightarrow} B \stackrel{T_2}{\rightarrow} C$, we are assuming that 
firing of $C$ after $T_2$ given that $B$ has fired is independent of $A$ having 
fired earlier. That is why we have used $p=\rho_A (e_0)^2$ while calculating our 
threshold. But suppose we do not want to assume this. Then we can have a null hypothesis 
that is specified by bounds on different conditional probabilities. Suppose 
$e_2(x,y,T)$ is the conditional probability that $y$ fires after $T$ given $x$ has 
fired and suppose $e_3(x,y,z,T_1,T_2)$ be the probability that $y$ fires after $T_1$ and 
$z$ fires after another $T_2$ given $x$ has fired. Now we specify the null hypothesis 
in terms of two parameters as: $e_2(x,y,T) < e_{02}, \ \forall x,y$ and 
$e_3(x,y,z,T_1,T_2) < e_{03}, \ \forall x,y,z$. Now for assessing significance of 
3-node episodes we can use $p=\rho_A e_{03}$. Our method of analysis is still 
applicable without any modifications. Of course, now the user has to specify two 
bounds on different conditional probabilities and he has to have some reasons for 
distinguishing between the two conditional probabilities. But the main point here 
is that the model is fairly general and can accommodate many such extensions. 

There are many other ways in which the idea presented here can be extended. Suppose 
we want to assess significance of synchronous firing patterns rather than sequential 
patterns based on the count of number of non-overlapped occurrences of the synchronous 
firing pattern. One possibility would be to use conditional probabilities of $A$ firing within an 
appropriate short time interval from $B$ in our null hypothesis and then use an 
appropriate expression for $p$ in our model. 
Another example could be that of analyzing occurrences of neuronal firing sequences 
that respect a pre-set order on the neurons as discussed in \cite{SS2006}. Suppose 
we want to assess the significance of count of such patterns of a fixed length. 
If we use our type of non-overlapped occurrences count as the statistic, then the 
model presented here can be used to assess the significance. Now the parameter $p$ would 
be the probability of occurrence of a sequence  of that length (which respects the 
global order on the neurons) starting from any time instant. 
 For a given null hypothesis, e.g., of independence, this 
would be a combinatorial problem similar to the one tackled in \cite{SS2006}. Once we 
can derive an expression for $p$ we can use our method for assessing significance. 

Though we did not discuss the computational issues in this paper, the data mining algorithms 
used for discovering sequential patterns are computationally efficient (see \cite{PSU2008} 
for details). One computational issue that may be relevant for this paper may be that of data  
sufficiency. All the results reported here are on spike data of 20 sec duration with 
background spiking rate of 20 Hz. (That works out to about 400 spikes per neuron on the 
average in the data). From fig.~\ref{fig:acc-3-4-5} we can see that, with this much of data,  
we can certainly distinguish between connection strengths that differ by about 0.2 on the 
conditional probability scale. (Notice that, in the figure, the mean plus three sigma range 
of the count distribution at a connection strength is below our threshold (with $k=3$) at a 
connection strength 0.2 more). In fig.~\ref{fig:rankorder-3-4-5} we showed that we can 
reliably rank order connection strengths with about the same resolution. Thus we can say that 
20 sec of data is good enough for this level of discrimination. Obviously, if we need to distinguish 
between only widely different strengths, much less data would suffice. 
 
In terms of computational issues, we feel that one of the important conclusions from this paper  
 is that temporal data mining may be an attractive approach for tackling the 
problem of discovering firing patterns (or microcircuits) in multi-neuronal spike trains.   
In temporal data mining literature, episodes are, in general, partially ordered sets of 
event types. Here we used the methods for discovery of serial episodes which correspond to 
our sequential patterns. A general episode would correspond to a graph of interconnections 
among neurons. However, at present, there are no efficient algorithms for discovering 
frequently occurring graph patterns from a data stream of events. Extending our data mining 
algorithm and our analysis technique to tackle such graph patterns is another interesting 
open problem. This would allow for discovery of more general microcircuits from 
 spike trains. 

In summary, we feel that the general approach presented here 
 has a lot of potential and it can 
be specialized to handle many of the data analysis needs in multi-neuronal spike 
train data. We would be exploring many of these issues in our future work.

\begin{center}
{\bf Acknowledgments} \\
\end{center}

We wish to thank Mr. Debprakash Patnaik and Mr. Casey Diekman for their help 
in preparing this paper. The simulator described here as well as the data mining package 
for analyzing data streams is written by 
Mr. Patnaik  \cite{PSU2008} and he has helped in running 
the simulator. Mr. Diekman has helped in generating  all the 
figures. The work reported here is partially supported by a project funded by 
General Motors R\&D Center, Warren through SID, Indian Institute of Science, 
Bangalore.

\end{document}